\documentclass[11pt]{article}
\renewcommand{\theequation}{\arabic{section}.\arabic{equation}}

\newtheorem{example}{Example}[section]
\newtheorem{theorem}{Theorem}[section]
\newtheorem{corollary}{Corollary}[section]
\usepackage{bm}
\usepackage{amssymb}
\usepackage{amsmath}
\usepackage{epsfig}

\begin{document}
\begin{center}
{\LARGE\bf The chain rule for functionals}\\[1ex]
{\LARGE\bf with applications to functions of moments}\\[1ex]
by\\[1ex]
Christopher S. Withers\\
Applied Mathematics Group\\
Industrial Research Limited\\
Lower Hutt, NEW ZEALAND\\[2ex]
Saralees Nadarajah\\
School of Mathematics\\
University of Manchester\\
Manchester M13 9PL, UK
\end{center}
\vspace{1.5cm}
{\bf Abstract:}~~The chain rule for derivatives of a function of a function is extended to
a function of a statistical functional, and applied to obtain
approximations to the cumulants, distribution and quantiles of functions
of sample moments, and so to obtain third order confidence intervals
and estimates of reduced bias for functions of moments.
As an example we give the distribution of the standardized skewness for
a normal sample to magnitude $O(n^{-2})$, where $n$ is the sample size.



\section{Introduction}
\setcounter{equation}{0}

The derivatives introduced by von Mises (1947) and their subsequent versions have
wide ranging applications in statistics.
Two prominent application areas are the
construction of nonparametric confidence intervals and analytic bias reduction.

Suppose we want to construct a nonparametric confidence interval of level $\alpha + O(n^{-3/2})$
for a smooth functional $T(F)$ based on $\widehat{F}$ say, the sample or empirical
distribution for a sample of size $n$ from $F$.
Withers (1983) showed that the limit can be given in terms
of integrals of products of von Mises derivatives evaluated at  $\widehat{F}$.
First one Studentizes using the asymptotic
variance of $n^{1/2}\{T(\widehat{F})-T(F)\}$,
$a_{21}({F})=[1^2]_T = \int_{-\infty}^\infty T_F(x_1)^2dF(x_1)$,
where $T_F(x)$ is the first derivative or influence function of $T(F)$.

Similarly, it is known that for smooth $T(F)$,  an estimate of $T(F)$ of bias $O(n^{-2})$ is
$T(\widehat{F})-n^{-1}a_{11}(\widehat{F})$, where
$a_{11}({F})=[11]_T = \int_{-\infty}^\infty T_F(x_1,x_1)dF(x_1)$ and  $T_F(x_1,x_2)$ is the second derivative of $T(F)$.
For convenience we refer to these  integrals of products of derivatives
like $[1^2]_T$ and $[11]_T$ as {\it bracket functions}.

Other recent application areas of von Mises derivatives include:
least squares support vector regression filtering methods,
bootstrapping,
functional principal components analysis,
linearization and composite estimation,
dimension reduction,
quantile regressions,
machine learning,
cusum statistics,
methods of sieves and penalization,
change point estimation,
Hadamard differentiability,
change-of-variance function,
measuring and testing dependence by correlation of distances,
empirical finite-time ruin probabilities (Loisel {\it et al.}, 2009),
nonparametric maximum likelihood estimators (Nickl, 2007),
estimating mean dimensionality of analysis of variance decompositions,
monotonicity of information in the Central Limit Theorem,
generalizations of the Anderson-Darling statistic,
$M$-estimation,
$U$-statistics (Volodko, 2011),
information criteria in model selection,
goodness-of-fit tests for kernel regression,
empirical Bayes estimation, and estimation of Kendall's tau.

The aim of this paper is to develop tools for extending the use of von Mises derivatives.
In Section 2, we extend Faa di Bruno's chain rule for the derivative of a function
of a univariate function to functions of a multivariate function
and show how it can be applied to a function of a function of $F$,
say $T(F)=g(S(F))$, where $g:{\mathbb{R}}^a \rightarrow {\mathbb{R}}$
is a smooth function and $S(F)$ a smooth functional.

In Section 3, we apply it to obtain derivatives and bracket functions for
powers, products, quotients, standardized and Studentized functionals.

Section 4 gives the general derivative for a moment and applies previous results
to obtain expansions up to $O(n^{-2})$ for the distribution and
quantiles of functions of sample moments.
As an example we give the distribution of the standardized skewness for
a normal sample to magnitude $O(n^{-2})$, where $n$ is the sample size.
Also we give confidence intervals and bias reduction methods for functions
of moments.

Some of the results in the paper follow easily from Withers (1983, 1987), see Theorems 3.1 to 3.3.
But these results are not the main contributions of this paper.
The main contributions are: 1) the tools developed to compute
von Mises type derivatives, see Theorems 2.1 and 2.2;
2) their applications to obtain bracket functions for general functionals, see Examples 3.1 to 3.4 and Appendix A.
The functionals considered by these examples include $T (F) = g(S(F))$, where $g$ is a univariate function,
$T(F) = S_1 (F) S_2 (F)$, a product of two functionals,
Studentized forms of $T (F)$ and $T (F) = U(F) g(S(F))$, where $S (F)$ is real valued;
3) also the applications of Theorems 2.1 and 2.2 to obtain derivatives of central moments,
see Theorem 4.1, Corollary 4.1, Corollary 4.2 and Appendix B.

Fisher and Wishart gave unbiased estimates only
for cumulants and their products: see, for example, Stuart and Ord (1987).
Our two methods for bias reduction apply to
any smooth functional - and our second estimate reduces to their results for
the cases they consider.
Also our method does not need to use unbiased estimates of cumulants
to reduce the bias of functions of cumulants.

Analogous to Fisher's tables for his $k$-statistics and their cumulants,
Appendix B gives the terms needed for bias reduction of any smooth function
of one or more moments.

\section{Chain rules for functions and functionals}
\setcounter{equation}{0}

Let $s$ and $g$ be real functions on ${\mathbb{R}}$ with finite derivatives.
Comtet (1974, page 137) gives  Faa di Bruno's chain rule for the $r$th derivative of
\begin{eqnarray}
t(x)=g\left(s(x)\right)
\nonumber
\end{eqnarray}
for $r=1,2, \ldots$ in the form
\begin{eqnarray}
t^{(r)}(x)=\sum_{h=1}^r  g^{(h)} \left(s(x)\right) B_{rh} ({\bf s})
\label{xxx}
\end{eqnarray}
evaluated at ${\bf s} = (s_1, s_2, \ldots)$, $s_i=s^{(i)}(x)$, where $B_{rh}$ is the {\it partial exponential Bell polynomial}
defined by the coefficients in the formal expansion in powers of real $\varepsilon $,
\begin{eqnarray}
\left( \sum_{i=1}^\infty \varepsilon^i s_i/i! \right)^j/j!
= \sum_{r=j}^\infty \varepsilon^r {B}_{rj} ({\bf s})/r!
\nonumber
\end{eqnarray}
for $j \geq 0$.
Comtet (1974) shows they are given by
\begin{eqnarray}
{B}_{rj} ({\bf s}) / j!  =  \sum_{n \mbox{ in } {\mathbb{N}}^r}
\left\{ \frac {\displaystyle s_1^{n_1}\cdots  s_r^{n_r}}{\displaystyle n_1!\cdots n_r!} :
n_1+\cdots +n_r =j,
\
1 \cdot n_1+\cdots +r \cdot n_r =r \right\},
\nonumber
\end{eqnarray}
where ${\mathbb{N}}=\{0,1,2, \ldots \}$.
Comtet (1974, page 307) tables them for $r\leq 12$.
For example,
\begin{eqnarray}
&&
B_{r1}({\bf s}) = s_r,
\
B_{rr}({\bf s})=s_1^r,
\label{r1}
\\
&&
B_{32}({\bf s}) = 3s_1s_2,
\
B_{42}({\bf s})=4s_1s_3+3s_2^2,
\
B_{43}({\bf s})=6s_1^2s_2.
\label{r3}
\end{eqnarray}
Theorem 2.1 provides an extension of (\ref{xxx}) to the case  $s:{\mathbb{R}}^a\rightarrow {\mathbb{R}}^b$
and $g:{\mathbb{R}}^b\rightarrow {\mathbb{R}}$.

\begin{theorem}
Define the partial derivatives
\begin{eqnarray*}
&&
t_{ \cdot j_1 \cdots j_r}(x) = \partial^{r}t(x)/\partial x_{j_1} \cdots \partial x_{j_r},
\\
&&
s_{i \cdot j_1 \cdots j_r} = s_{i \cdot j_1 \cdots j_r}(x)  =
\partial^{r}s_{i}(x)/\partial x_{j_1} \cdots \partial x_{j_r},
\\
&&
g_{ \cdot j_1 \cdots j_r}(s) =  \partial^{r}g(s)/\partial s_{j_1} \cdots \partial s_{j_r}.
\end{eqnarray*}
The extension of (\ref{xxx}) is
\begin{eqnarray}
t_{ \cdot 1\cdots r}(x) =\sum^{r}_{h=1}g_{ \cdot i_{1} \cdots i_{h}}
\left(s(x)\right) B_{r}^{i_1\cdots i_h}({\bf s}).
\label{mult}
\end{eqnarray}
In (\ref{mult}) and throughout, we use the tensor sum convention that
repeated indices $i_1,i_2, \ldots$
are implicitly summed over their range ($1, \ldots, b$ in the case of (\ref{mult})).
\end{theorem}

Note that $B_{r}^{i_1\cdots i_h}({\bf s})$ can be written down on sight from $B_{rh}$.
Some particular cases of $B_{r}^{i_1\cdots i_h}({\bf s})$ can be obtained from (\ref{r1}) and (\ref{r3}):
\begin{eqnarray}
&&
B_{r}^i({\bf s}) = s_{i \cdot 1 \cdots r},
\ B_{r}^{i_1\cdots i_r}({\bf s})=s_{i_1\cdot 1} \cdots s_{i_r\cdot r},
\nonumber
\\
&&
B_{3}^{i_1 i_2 }({\bf s}) = \sum^3 s_{i_1 \cdot 1}s_{i_2 \cdot 2 3},
\nonumber
\\
&&
B_{4}^{i_1 i_2}({\bf s}) = \sum^{4} s_{i_{1}\cdot 1} s_{i_{2} \cdot 234}
+\sum^{3} s_{i_{1} \cdot 12}s_{i_{2} \cdot 34},
\
B_{4}^{i_1 i_2i_3}({\bf s}) = \sum^{6}s_{i_{1}\cdot 1}s_{i_{2}\cdot 2}s_{i_{3}\cdot 34},
\nonumber
\end{eqnarray}
where
\begin{eqnarray}
\sum^{r} h_{1\cdots r} = h_{1\cdots r}+h_{2 \cdots r1}+ \cdots +h_{r1 \cdots r-1}.
\nonumber
\end{eqnarray}
A form of the multivariate chain rule (\ref{mult}) was given in Withers (1984).

Let $\cal F$ be a convex set of probability measures on a measurable space $(\Omega, A)$.
Suppose for $x \in \Omega$, that $\delta_{x}$ lies in ${\cal F}$, where $\delta_{x}$ is
the measure putting mass $1$ at $x$ and $0$ elsewhere.
Let $x$, $\{ x_{i}\}$ be points in $\Omega$.
Let  $F$ lie in ${\cal F}$, and $T: F\rightarrow {\mathbb{R}}$ be some functional.
Define the $r$th derivative of $T(F)$ at $(x_{1}, \ldots, x_{r})$,
\begin{eqnarray*}
T_{\cdot 1\cdots r}=T_{F}\left(x_{1}\cdots x_{r}\right)=
T_{F}^{(r)} \left(x_{1}, \ldots, x_{r} \right),
\end{eqnarray*}
as in Withers (1983).
The only  derivative we need give here is the first, also known as the {\it influence function}:
\begin{eqnarray*}
T_{\cdot 1} \left( x_1 \right) = T_F\left(x_1\right)= \lim_{\epsilon\downarrow 0}
\left\{ T\left((1-\epsilon)F + \epsilon\delta_{x_1} \right)-T(F)\right\}/\epsilon.
\end{eqnarray*}
For example, $T(F) = \int_{-\infty}^\infty g(x)dF(x)$ has first derivative $T_{\cdot 1} = T_{\cdot 1} (x) = g(x)-T(F)$.
The results stated  in Withers (1983) for $\Omega = {\mathbb{R}}^{s}$ generalize
immediately to general $\Omega$.
In particular, the rule (\ref{2.6}) for the derivative
of the $r$th derivative may be stated as
\begin{eqnarray}
\left(T_{\cdot 1 \cdots r}\right)_{r+1}=T_{\cdot 1 \cdots r+1}  -
\sum_{i=1}^{r}\left[T_{\cdot 1 \cdots r+1}\right]_{i},
\label{2.1}
\end{eqnarray}
where $[T_{\cdot 1\cdots r+1}]_{i}=T_{\cdot 1\cdots r+1}$ with the $i$th argument dropped.
So,
\begin{eqnarray*}
&&
\left(T_{\cdot 1}\right)_{2}  =  T_{\cdot 12}-T_{\cdot 2},
\\
&&
\left(T_{\cdot 12}\right)_{3} = T_{\cdot 123}-T_{\cdot 23}-T_{\cdot 13}.
\end{eqnarray*}
In this way higher derivatives may be calculated from successive first derivatives.
For example, the second derivative of $\int_{-\infty}^\infty g (x) dF (x)$ is zero.
Now suppose for some function $g: {\mathbb{R}}^{b}\rightarrow {\mathbb{R}}$,
\begin{eqnarray}
T(F)=g\left(S(F)\right)\ \mbox{where}\ S(F)\ \mbox{is a real functional in}\ {\mathbb{R}}^{d}.
\label{2.2}
\end{eqnarray}
Applying (\ref{2.1}) gives
\begin{eqnarray}
&&
T_{\cdot 1} = g_{\cdot i}S_{i\cdot 1},
\label{2.3}
\\
&&
T_{\cdot 12} = g_{\cdot i}S_{i\cdot 12}+g_{\cdot ij}S_{i\cdot 1}S_{j\cdot 2},
\label{2.4}
\\
&&
T_{\cdot 123} = g_{\cdot i}S_{i\cdot 123}+g_{\cdot ij}\sum^{3} S_{i\cdot 1}S_{j\cdot 23}+g_{\cdot ijk}S_{i\cdot 1}
S_{j\cdot 2}S_{k\cdot 3},
\label{2.5}
\end{eqnarray}
and so on, where $S_{a\cdot 12 \cdots }$ is the $r$th derivative of  $S_a(F)$.
Despite the fact that by (\ref{2.1}) the derivative of a derivative is {\it not} a second
derivative, the  expressions (\ref{2.3})-(\ref{2.5}) are precisely those for the
derivatives of a function of a vector function of a vector given in  (\ref{mult}).
That is,
\begin{eqnarray}
T_{ \cdot 1\cdots r} =\sum^{r}_{h=1}g_{ \cdot i_{1} \cdots i_{h}}
\left(S(F)\right) B_{r}^{i_1\cdots i_h} ({\bf S}),
\label{multF}
\end{eqnarray}
where ${\bf S} = (S_1, S_2, \ldots)$, $S_i=S^{(i)}(F)$.
A proof that  (\ref{multF}) holds for general $r$ follows using  (\ref{2.1})
and induction.
The result can be formally stated as follows.

\begin{theorem}
If (\ref{2.2}) holds, $T_{\cdot 1 \cdots r}$ is given by the chain rule for
\begin{eqnarray*}
T_{\cdot 1 \cdots r} = T_{\cdot 1 \cdots r} (x) = \partial^{r} T(x)/\partial x_{1} \cdots \partial x_{r}
\nonumber
\end{eqnarray*}
when $T(x)=g(S(x))$ for
$S(x): {\mathbb{R}}^{r}\rightarrow {\mathbb{R}}^{d}$ with
$S_{i\cdot 1 \cdots r} = S_{i\cdot 1 \cdots r} (x) = \partial^{r}S_{i}(x)/\partial x_{1} \cdots \partial x_{r}$
re-interpreted as $S_{iF}(x_{1} \cdots x_{r})$ and $S(x)$ as $S(F)$.
So,
\begin{eqnarray}
T_{\cdot 1 \cdots r}=\sum_{h=1}^{r}g_{\cdot i_{1} \cdots i_{h}}\sum_{n}
\sum_{\Pi}^{m(n)}S_{i_{1} \cdot \Pi_{1}} \cdots S_{i_{h} \cdot \Pi_{h}},
\label{2.6}
\end{eqnarray}
where $g_{\cdot i_{1} \cdots i_{h}} = g_{\cdot i_{1} \cdots i_{h}} (y) = \partial_{i_{1}} \cdots \partial_{i_{h}}g(y)$ at $y=S(F)$
for $\partial_{i}=\partial/\partial y_{i}$,
$\sum_{n}$ sums over $n=(n_{1} \cdots n_{r}) \in {\mathbb{N}}^{r}$ satisfying
$\sum_{i=1}^r n_{i}=k$, $\sum_{i = 1}^r in_{i}=r$, $m(n)=r!/\prod^{r}_{i=1} i!^{n_{i}}n_{i}!$,
the partition function, and
$\sum_{\Pi}^{m(n)}$ sums over all partitions $(\Pi_{1} \cdots \Pi_{k})$ of
$(1 \cdots r)$ with $i\ \Pi$'s of length $n_{i}$.
\end{theorem}

Corollary 2.1 applies Theorem 2.2 to obtain the next two derivatives.

\begin{corollary}
We
\begin{eqnarray}
T_{\cdot 1234}
&=&
g_{\cdot i} S_{i \cdot 1234}+g_{\cdot i_{1}i_{2}}\left(\sum^{4}S_{i_{1} \cdot 1}S_{i_{2} \cdot 234}
+\sum^{3}S_{i_{1} \cdot 12}S_{i_{2} \cdot 34}\right)
\nonumber
\\
&&
+ g_{\cdot i_{1}i_{2}i_{3}}\sum^{6}S_{i_{1} \cdot 1}S_{i_{2} \cdot 2}
S_{i_{3} \cdot 34}+g_{\cdot i_{1} \cdots i_{4}}S_{i_{1} \cdot 1} S_{i_{2} \cdot 2}S_{i_{3} \cdot 3}
S_{i_{4} \cdot 4},
\nonumber
\\
T_{\cdot 12345}
&=&
g_{\cdot i}S_{i \cdot 1 \cdots 5}+g_{\cdot i_{1}i_{2}}\left(\sum^{4}S_{i_{1} \cdot 1}
S_{i_{2} \cdot 2345}+\sum^{10}S_{i_{1} \cdot 12}S_{i_{2} \cdot 345}\right)
\nonumber
\\
&&
+g_{\cdot i_{1}i_{2}i_{3}}\left(\sum^{10}S_{i_{1} \cdot 1}S_{i_{2} \cdot 2}
S_{i_{3} \cdot 345}+\sum^{15}S_{i_{1} \cdot 1}S_{i_{2} \cdot 23}S_{i_{3} \cdot 45}\right)
\nonumber
\\
&&
+g_{\cdot i_{1} \cdots i_{4}}\sum^{10}S_{i_{1} \cdot 1}S_{i_{2} \cdot 2}S_{i_{3} \cdot 3}
S_{i_{4} \cdot 45}+g_{\cdot i_{1} \cdots i_{5}}S_{i_{1} \cdot 1} \cdots S_{i_{5} \cdot 5},
\nonumber
\end{eqnarray}
so
\begin{eqnarray*}
\left[1^{k}\right]_T
&=&
g_{\cdot i_1}\cdots g_{\cdot i_k}
\left[ S_{i_1 \cdot 1}\cdots S_{i_k \cdot 1}\right],
\\
\left[ 11 \right]_T
&=&
g_{\cdot i}[{11}]_{S_i} +g_{\cdot i}g_{\cdot j}
\left[S_{i \cdot 1} S_{j \cdot 1}\right],
\\
\left[ 1,2,12 \right]_T
&=&
g_{\cdot i_1} g_{\cdot i_2} g_{\cdot i_3}
\left[S_{i_1 \cdot 1} S_{i_2 \cdot 2} S_{i_3 \cdot 12}\right]
+ g_{\cdot i_1} g_{\cdot i_2} g_{\cdot i_3 i_4}
\left[S_{i_1 \cdot 1} S_{i_3 \cdot 1}\right]
\left[S_{i_2 \cdot 1} S_{i_4 \cdot 1}\right],
\\
\left[ 111 \right]_T
&=&
g_{\cdot i}[{111}]_{S_i}
+3g_{\cdot ij} \left[S_{i \cdot 1} S_{j \cdot 11}\right]
+g_{\cdot ijk}\sum^3 \left[S_{i \cdot 1} S_{j \cdot 1} S_{k \cdot 1}\right],
\\
\left[ 1122 \right]_T
&=&
g_{\cdot i}[{1122}]_{S_i}
+g_{\cdot i_{1}i_{2}}\left({4}\left[S_{i_{1} \cdot 1}S_{i_{2} \cdot 122}\right]
+2\left[S_{i_{1} \cdot 12}S_{i_{2} \cdot 12}\right]+[11]_{S_i}[11]_{S_j} \right)
\nonumber
\\
&&
+ 2g_{\cdot i_{1}i_{2}i_{3}}
\left(2 \left[S_{i_{1} \cdot 1}S_{i_{2} \cdot 2} S_{i_{3} \cdot 12}\right]
+\left[S_{i_1\cdot 1}S_{i_2\cdot 1} \right] [11]_{S_{i_3}}\right)
\nonumber
\\
&&
+g_{\cdot i_{1} \cdots i_{4}}
\left[S_{i_{1} \cdot 1} S_{i_{2} \cdot 1}\right]
\left[S_{i_{3} \cdot 1}         S_{i_{4} \cdot 1}\right],
\\
\left[ 1,122 \right]_T
&=&
g_{\cdot i} g_{\cdot j} \left[S_{i \cdot 1} S_{j \cdot 122}\right]
+g_{\cdot i}g_{\cdot jk}\left( \left[S_{i \cdot 1} S_{j \cdot 1}\right] [11]_{S_k}
+2\left[S_{i \cdot 1}S_{j \cdot 2} S_{k \cdot 12}\right] \right)
\\
&&
+g_{\cdot i}g_{\cdot jkl}
\left[S_{i \cdot 1} S_{j \cdot 1}\right]
\left[S_{k \cdot 1} S_{l \cdot 1}\right],
\\
\left[12^2\right]_T
&=&
g_{\cdot i} g_{\cdot j} \left[S_{i \cdot 12} S_{j \cdot 12}\right]
+2g_{\cdot i}g_{\cdot jk} \left[S_{i \cdot 12} S_{j \cdot 1}S_{k \cdot 2}\right]
+g_{\cdot ij}g_{\cdot kl}
\left[S_{i \cdot 1} S_{k \cdot 1}\right]
\left[S_{j \cdot 1} S_{l \cdot 1}\right],
\end{eqnarray*}
and so on, where
\begin{eqnarray*}
&&
\left[S_{i\cdot 1} S_{j\cdot 1}\right] = \int_{-\infty}^\infty S_{i\cdot 1} S_{j\cdot 1} dF\left(x_1\right),
\\
&&
\left[S_{i\cdot 1} S_{j\cdot 1} S_{k\cdot 12}\right] =
\int_{-\infty}^\infty \int_{-\infty}^\infty
S_{i\cdot 1} S_{j\cdot 1} S_{k\cdot 12} dF\left(x_1\right) dF\left(x_2\right),
\end{eqnarray*}
and so on.
\end{corollary}

\section{Some applications}
\setcounter{equation}{0}

Let $\widehat{F}$ be the empirical distribution of a random sample of size $n$ from $F$.
By Withers (1983), for a broad class of $T$, the cumulants of $T(\widehat{F})$ satisfy
\begin{eqnarray}
\kappa_{r}\left(T \left( \widehat{F} \right) \right)
\approx \sum_{i=r-1}^{\infty}n^{-i}a_{ri}
\nonumber
\end{eqnarray}
for $r \geq 1$, where the {\it cumulant coefficient}
$a_{ri}(T)=a_{ri}$ is a certain function of the derivatives of $T(F)$.
The most important are $a_{10}=T(F)$,
\begin{eqnarray}
&&
a_{21} = \left[1^{2}\right]_T = \left[T_{\cdot 1}^{2}\right],
\label{3.2}
\\
&&
a_{11}  =  [11]_T/2 = \left[T_{\cdot 11}\right]/2,
\label{3.3}
\\
&&
a_{32}  =  \left[1^{3}\right]_T+3[1,2,12]_T  =
\left[T_{\cdot 1}^{3}\right] + 3\left[T_{\cdot 1}T_{\cdot 2}  T_{\cdot 12}\right],
\label{3.4}
\end{eqnarray}
where
\begin{eqnarray*}
&&
\left[f \left( T_{\cdot 1},T_{\cdot 11}, \ldots \right)\right] =
\int_{-\infty}^\infty f \left( T_{\cdot 1}, T_{\cdot 11}, \ldots \right)dF_{1} \left( x_1 \right),
\\
&&
\left[f\left( T_{\cdot 1}, T_{\cdot 2}, T_{\cdot 11},
T_{\cdot 12}, T_{\cdot 22}, T_{\cdot 122}, \ldots \right) \right]
\\
&&
\qquad \qquad \qquad =
\int_{-\infty}^\infty \int_{-\infty}^\infty f\left(T_{\cdot 1}, T_{\cdot 2},
T_{\cdot 11}, T_{\cdot 12}, T_{\cdot 22}, T_{\cdot 122}, \ldots \right)
dF_{1} \left( x_1 \right) dF_{2} \left( x_2 \right),
\end{eqnarray*}
and so on, for $F_{i}=F(x_{i})$, and
\begin{eqnarray*}
\left[1^i,2^j,11^k,12^l,22^m, \ldots \right]_T =
\left[ T_{\cdot 1}^i, T_{\cdot 2}^j, T_{\cdot 11}^k, T_{\cdot 12}^l, T_{\cdot 22}^m, \ldots \right],
\end{eqnarray*}
and so on.
We refer to the functionals $[\cdots]$ as {\it bracket functions}.
They are the building blocks for the cumulant coefficients
$a_{ri}$ and the  cumulant coefficients of the Studentized statistics,
and hence for the Edgeworth-Cornish-Fisher expansions of the standardized form of $T(\widehat{F})$,
\begin{eqnarray}
Y_n = \left(n/ a_{21}\right)^{1/2}\left\{T \left( \widehat{F} \right) - T(F)\right\},
\label{Y_n}
\end{eqnarray}
and its Studentized form.
They are also the building blocks for obtaining nonparametric confidence
intervals and estimates of low bias for $T(F)$.

As a start we have these approximations to the bias, variance, and skewness of $T(\widehat{F})$:
\begin{eqnarray*}
&&
\mathbb{E} \left[ T\left(\widehat{F}\right) \right] = T(F)+n^{-1}a_{11}+O\left(n^{-2}\right),
\nonumber
\\
&&
\mbox{var} \left[ T\left(\widehat{F}\right) \right] = n^{-1}a_{21}+O\left(n^{-2}\right),
\nonumber
\\
&&
\mu_{3}\left[T\left(\widehat{F}\right) \right] = n^{-2}a_{32}+O\left(n^{-3}\right),
\nonumber
\end{eqnarray*}
where $\mu_{3}[X]  = \mathbb{E} [(X - \mathbb{E} [X])^{3}]$.

Theorem 3.1 lists the bracket functions needed for bias and bias reduction.
Theorem 3.2 lists the bracket functions needed
for Edgeworth-Cornish-Fisher expansions.
Theorem 3.3 lists the bracket functions needed
for nonparametric confidence intervals.

\begin{theorem}
Under regularity conditions,
\begin{eqnarray*}
&&
\mathbb{E} \left[ T\left(\widehat{F}\right) \right] = T(F)+\sum_{i=1}^{j}n^{-i}a_{1i}+O\left(n^{-j-1}\right),
\\
&&
a_{11}  =  [11]_T/2,
\\
&&
a_{12}  = [111]_T/6+[1122]_T/8,
\\
&&
a_{13} = [11 11]_T/24+[1122]_T/12+[112233]_T/48,
\end{eqnarray*}
and so on.
The estimates of $T(F)$ of bias $O(n^{-j-1})$ are
\begin{eqnarray}
T\left(\widehat{F}\right)+\sum_{i = 1}^{j}n^{-i}T_{i}\left(\widehat{F}\right)
\mbox{ and }
T\left(\widehat{F}\right)+\sum_{i = 1}^{j}S_{i}\left(\widehat{F}\right)/(n-1)_{i},
\label{3.7}
\end{eqnarray}
where $(m)_{i} = m!/(m-i)!=m(m-1) \cdots (m-i+1)$ and
\begin{eqnarray}
&&
T_{ 1}(F) = S_{1}(F) =-[11]_T/2,
\label{3.8}
\\
&&
T_{ 2}(F) = [111]_T/3+[11 22]_T/8-[11]_T/2,
\label{3.9}
\\
&&
T_{ 3}(F) = -[11]_T/2+[111]_T-[11 11]_T/4+3[11 22]_T/4 -[11122]_T/6-[11 22 33]_T/48,
\nonumber
\\
&&
S_{2}(F) = [111]_T/3+[1122]_T/8,
\nonumber
\\
&&
S_{3}(F) = -[11 11]_T/4 +3[1122]_T/8-[11122]_T/6-[112233]_T/48.
\nonumber
\end{eqnarray}
\end{theorem}

\noindent
{\bf Proof:}~~Follows by equation (2.4) of Withers (1987).
\
$\Box$

\begin{theorem}
The `reduced' Edgeworth and Cornish-Fisher expansions
of $T(\widehat{F})$ to $O($ $n^{-(j+1)/2}$ $)$ needs
\begin{eqnarray}
&&
\mbox{for}\ j = 0: a_{21},
\nonumber
\\
&&
\mbox{for}\ j = 1: a_{11}\ \mbox{and}\ a_{32},
\nonumber
\\
&&
\mbox{for}\ j = 2: a_{22}=[1,11]_T+\left[12^{2}\right]_T/2+[1,122]_T,
\nonumber
\\
&&
\ a_{43} = \left[1^{4}\right]_T-3\left[1^{2}\right]_T^{2}+12\left[1,2^{2},12\right]_T+
12[1,2,13,23]_T +4[1,2,3,123]_T,
\label{3.16}
\end{eqnarray}
and so on.
In particular, for $Y_n$ of (\ref{Y_n}), under regularity conditions,
\begin{eqnarray*}
P\left(Y_n\leq x\right) = \Phi(x)-\phi(x)\left[ n^{-1/2}h_1(x)+n^{-1}h_2(x) \right] +O\left(n^{-3/2}\right)
\end{eqnarray*}
for $h_1 = A_{11}+A_{32}He_2/6$ and
$h_1 = (A_{22}+A_{11}^2)He_1/2 +(A_{43}+4A_{11}A_{32})He_3/24+A_{32}^2He_5/72$,
where $\Phi$, $\phi$ are the distribution and density of a unit normal random
variable, $He_r$ is the $r$th Hermite polynomial, and
$A_{ri}=a_{ri}/a_{21}^{r/2}$, the standardized cumulant coefficient.
\end{theorem}

\noindent
{\bf Proof:}~~Follows by Withers (1983).
\
$\Box$

The regularity conditions needed for Theorems 3.1 and 3.2 are the same as those given in Withers (1983, 1987).
So, they are not stated here.

\begin{theorem}
A confidence interval
for $T(F)$ of level $1-\alpha +O(n^{-(j+1)/2})$ requires the bracket functions
\begin{eqnarray}
&&
\mbox{for}\ j  =  0: a_{21}=\left[1^{2}\right]_T,
\nonumber
\\
&&
\mbox{for}\ j = 1: [11]_T,
\
\left[1^{3}\right]_T,
\
[1,2,12]_T,
\nonumber
\\
&&
\mbox{for}\ j = 2: [1,11]_T,
\left[12^{2}\right]_T, [1,122]_T,
\left[1^{4}\right]_T,
\left[1,2^{2},12\right]_T,
\nonumber
\\
&&
\qquad \qquad \qquad \qquad
[1,2,13,23]_T,
[1,2,3,123]_T.
\label{3.17}
\end{eqnarray}
\end{theorem}

\noindent
{\bf Proof:}~~Follows by Theorem 5.1 in Withers (1983).
\
$\Box$

By Withers (1989), the bracket functions in Theorem 3.3
are also the terms needed for the distribution
and quantiles of the Studentized form of $T(F)$ to $O(n^{-(j+1)/2})$.

For the distribution of $\mid T(\widehat{F})-T(F)\mid$ to $O(n^{-j-1})$ or for
a symmetric confidence interval for $T(F)$ of level $1-\alpha +O(n^{-j-1})$ one
needs, by equations (2.4) and (2.5) of Withers (1982), $a_{21}$ for $j = 0$
and  $a_{11}$, $a_{32}$, $a_{22}$, $a_{43}$ for $j = 1$.

For convenience, set $T=T(F)$ and $g_{i}=g^{(i)}(S(F))$ for $S(F)$ in ${\mathbb{R}}$.
\begin{example}
This example gives bracket functions for a function of a univariate functional.
Suppose (\ref{2.2}) holds with $b=1$.
Then
\begin{eqnarray*}
T_{\cdot 1}
&=&
g_{1} S_{\cdot 1},
\\
T_{\cdot 12}
&=&
g_{1}S_{\cdot 12}+g_{2}S_{\cdot 1}S_{\cdot 2},
\\
T_{\cdot 123}
&=&
g_{1}S_{\cdot 123}+g_{2} \sum^{3} S_{\cdot 1}S_{\cdot 23} +g_{3}S_{\cdot 1}S_{\cdot 2}S_{\cdot 3},
\\
T_{\cdot 1234}
&=&
g_{1}S_{\cdot 1234}+g_{2}\left(\sum^{4} S_{\cdot 1}S_{\cdot 234}+\sum^{3} S_{\cdot 12}S_{\cdot 34}\right)
\\
&&
+g_{3}\sum^{6}S_{\cdot 1}S_{\cdot 2}S_{\cdot 34}+g_{4}S_{\cdot 1}S_{\cdot 2}S_{\cdot 3}S_{\cdot 4},
\\
T_{\cdot 12345}
&=&
g_{1}S_{\cdot 1 \cdots 5}+g_{2}
\left(\sum^{4}S_{\cdot 1}S_{\cdot 2345} +\sum^{10}S_{\cdot 12}S_{\cdot 345}\right)
\\
&&
+g_{3}\left(\sum^{10}S_{\cdot 1}S_{\cdot 2}S_{\cdot 345}+
\sum^{15}S_{\cdot 1}S_{\cdot 23}S_{\cdot 45}\right)+g_{4}
\sum^{10}S_{\cdot 1}S_{\cdot 2}S_{\cdot 3}S_{\cdot 45}+g_{5}S_{\cdot 1} \cdots S_{\cdot 5}.
\end{eqnarray*}
So,
\begin{eqnarray*}
&&
\left[1^{k}\right]_T  = g_{1}^{k}\left[{1}^{k}\right]_S,
\\
&&
[11]_T  = g_{1}[{11}]_S+g_{2}\left[{1}^{2}\right]_S,
\\
&&
[1,2,12]_T =
g_{1}^{3}[{1},{2},{12}]_S+g_{1}^{2}g_{2}\left[{1}^{2}\right]_S^{2},
\\
&&
[111]_T =
g_{1}[{111}]_S+3g_{2}[{1},{11}]_S+g_{3}\left[{1}^{3}\right]_S,
\\
&&
[1122]_T =
g_{1}[{1122}]_S+g_{2}\left( 4[{1},{122}]_S+[{11}]_S^{2}+2\left[{12}^{2}\right]_S\right)
\\
&&
\qquad
+2g_{3}\left( \left[{1}^{2}\right]_S[{11}]_S+2[{1},{2},{12}]_S\right)
+g_{4}\left[{1}^{2}\right]_S^{2},
\\
&&
[1,122]_T =
g_{1}^2[1,{122}]_S+g_1g_{2}\left( \left[{1}^2\right]_S[{111}]_S+2[1,2,{12}]_S\right)
+g_1g_3\left[{1}^{2}\right]_S^2,
\\
&&
\left[12^2\right]_T  =
g_{1}^2\left[{12}^2\right]_S+2g_1g_2 \left[{1},2,{12}\right]_S+g_2^2\left[{1}^{2}\right]_S^2.
\end{eqnarray*}
\end{example}

\begin{example}
This example gives bracket functions for a product.
Suppose that $T(F)  =  S_{1}(F)S_{2}(F)$.
Then
\begin{eqnarray*}
T_{\cdot 1}
&=&
S_{2}S_{1 \cdot 1}+S_{1}S_{2 \cdot 1}=\sum^{\left(2\right)}
S_{1}S_{1 \cdot 1}\ \mbox{say,}
\nonumber
\\
T_{\cdot 12}
&=&
\left(S_{2}S_{1 \cdot 12}+S_{1}S_{2 \cdot 12}\right)+\left(S_{1 \cdot 1}S_{2 \cdot 2}+S_{2 \cdot 1}S_{1 \cdot 2}\right)
\\
&=&
\sum^{\left(2\right)}\left(S_{2}S_{1 \cdot 12}+S_{1}S_{2 \cdot 12}\right)\ \mbox{say,}
\nonumber
\\
T_{\cdot 123}
&=&
\left(S_{2}S_{1 \cdot 123}+S_{1}S_{2 \cdot 123}\right)+\sum^{3}\left(S_{1 \cdot 1}S_{2 \cdot 23}+
S_{2 \cdot 1}S_{1 \cdot 23}\right)
\nonumber
\\
&=&
\sum^{\left(2\right)}S_{2}S_{1 \cdot 123}+\sum^{\left(6\right)}S_{1 \cdot 1}S_{2 \cdot 23}\ \mbox{say,}
\nonumber
\\
T_{\cdot 1234}
&=&
\left(S_{2}S_{1 \cdot 1234}+S_{1}S_{2 \cdot 1234}\right)+\sum^{4}\left(S_{1 \cdot 1}S_{2 \cdot 234}+
S_{2 \cdot 1}S_{1 \cdot 234}\right)
\nonumber
\\
&&
+ \sum^{3}\left(S_{1 \cdot 12}S_{2 \cdot 34}+S_{2 \cdot 12}S_{1 \cdot 34}\right)
\nonumber
\\
&=&
\sum^{\left(2\right)}S_{2}S_{1 \cdot 1234}+\sum^{\left(8\right)}S_{1 \cdot 1}S_{2 \cdot 234}+
\sum^{6}S_{1 \cdot 2} S_{2 \cdot 34}\ \mbox{say.}
\end{eqnarray*}
So,
\begin{eqnarray*}
\left[1^{2}\right]_T
&=&
\sum^{\left(2\right)}S_{2}^{2}\left[{1}^{2}\right]_{S_1}
+2S_{1}S_{2}\left[S_{1 \cdot 1}S_{2 \cdot 1}\right],
\nonumber
\\
\left[11\right]_T
&=&
\sum^{\left(2\right)}S_{2}\left[11\right]_{S_1}+2\left[S_{1 \cdot 1}S_{2 \cdot 1}\right],
\nonumber
\\
\left[1^{3}\right]_T
&=&
\sum^{\left(2\right)}\left(S_{2}^{3}\left[ 1^{3} \right]_{S_1} + 3S_{2}^{2}\left[S_{2 \cdot 1}
S_{1 \cdot 1}^{2}\right]\right),
\nonumber
\\
\left[1,2,12\right]_T
&=&
\sum^{\left(2\right)}\Bigg\{ S_{2}^{3}\left[{1,2,12}\right]_{S_1}+S_{2}^{2}S_{1}
\Bigg(\left[S_{1 \cdot 1}S_{1 \cdot 2}S_{2 \cdot 12}\right]
\nonumber
\\
&&
+ 2\left[S_{1 \cdot 1}S_{2 \cdot 2}S_{1 \cdot 12}\right]\Bigg)+
2S_{2}^{2}\left[{1}^{2}\right]_{S_1}\left[S_{1 \cdot 1}S_{2 \cdot 1}\right]\Bigg\}
\nonumber
\\
&&
+ 2S_{1}S_{2}\left(\left[S_{1 \cdot 1}S_{2 \cdot 1}\right]^{2}+
\left[1^{2}\right]_{S_1}\left[1^{2}\right]_{S_2}\right),
\nonumber
\\
\left[111\right]_T
&=&
\sum^{\left(2\right)}\left(S_{2}\left[{111}\right]_{S_1}+3\left[S_{2 \cdot 1}S_{1 \cdot 11}\right]\right),
\nonumber
\\
\left[1122\right]_T
&=&
\sum^{\left(2\right)}\left(S_{2}\left[{1122}\right]_{S_1}+4\left[S_{1 \cdot 1}S_{2 \cdot 122}\right]\right)
+2\left[{11}\right]_{S_1}\left[{11}\right]_{S_2}+4\left[S_{1 \cdot 12}S_{2 \cdot 12}\right].
\nonumber
\end{eqnarray*}
\end{example}

\begin{example}
This example gives bracket functions for a Studentized function.
The  Studentized form of $T(F)$ is
\begin{eqnarray*}
T_0\left(\widehat{F}\right) = V\left(\widehat{F}\right)^{-1/2}
\left\{ T\left(\widehat{F}\right) - T(F) \right\}
\end{eqnarray*}
for $V(F)=a_{21}$.
Its bracket functions $[\cdots]_{T_0}$ (and so also its cumulant coefficients)
may be expressed in terms of the  bracket functions $[\cdots]_{T}$.
For details, see Appendix A of Withers (1989).
\end{example}

If one makes other assumptions such as symmetry of $F$ or a parametric form
for $F$, then $V(F)=a_{21}$ will generally  take a simpler form.
Similarly, in some circumstances one is interested in standardizing a functional
in a different way, for example, replacing $\mu_r$ by $\mu_r/\mu_2^{r/2}$.
The next example covers this situation for the special case of a $T(F)$
a function of a univariate functional.

\begin{example}
Suppose that $T(F)=  U(F)g(S(F))$ with $S(F)$ in ${\mathbb{R}}$.
Then
\begin{eqnarray*}
T_{\cdot 1}
&=&
g_{0}U_{\cdot 1}+g_{1}US_{\cdot 1},
\nonumber
\\
T_{\cdot 12}
&=&
g_{0}U_{\cdot 12}+g_{1}\left(US_{\cdot 12}+\sum^{2}U_{\cdot 1}S_{\cdot 2}\right)+g_{2}US_{\cdot 1}S_{\cdot 2},
\nonumber
\\
T_{\cdot 123}
&=&
g_{0}U_{\cdot 123}+g_{1}\left(US_{\cdot 123}+\sum^{\left(6\right)}U_{\cdot 1}S_{\cdot 23}\right)
\nonumber
\\
&&
+ g_{2}\sum^{3}\left(US_{\cdot 1}S_{\cdot 23}+U_{\cdot 1}S_{\cdot 2}S_{\cdot 3}\right)+
g_{3}US_{\cdot 1}S_{\cdot 2}S_{\cdot 3},
\nonumber
\\
T_{\cdot 1234}
&=&
g_{0}U_{\cdot 1234}+g_{1}\left(US_{\cdot 1234}+
\sum^{\left(8\right)}U_{\cdot 1}S_{\cdot 234}+\sum^{6}
U_{\cdot 12}S_{\cdot 34}\right)
\nonumber
\\
&&
+ g_{2}\left(U\sum^{4}S_{\cdot 1}S_{\cdot 234}+U\sum^{3}S_{\cdot 12}S_{\cdot 34}+\sum^{12}U_{\cdot 1}
S_{\cdot 2}S_{\cdot 34}+\sum^{6}U_{\cdot 12}S_{\cdot 3}S_{\cdot 4}\right)
\nonumber
\\
&&
+g_{3}\left(U\sum^{6}S_{\cdot 1}S_{\cdot 2}S_{\cdot 34}+
\sum^{4}U_{\cdot 1}S_{\cdot 2}S_{\cdot 3}S_{\cdot 4}\right)+g_{4}US_{\cdot 1}
S_{\cdot 2}S_{\cdot 3}S_{\cdot 4}.
\nonumber
\end{eqnarray*}
So, the cumulant coefficients $a_{21}$, $a_{11}$, $a_{32}$
needed for third order inference are given by
(\ref{3.2})-(\ref{3.4}) in terms of the bracket functions
\begin{eqnarray*}
\left[1^{2}\right]_T
&=&
g_{0}^{2}\left[{1}^{2}\right]_U+2g_{0}g_{1}U\left[U_{\cdot 1}S_{\cdot 1}\right]+
g_{1}^{2}U^{2}\left[{1}^{2}\right]_S,
\nonumber
\\
\left[11\right]_T
&=&
g_{0}\left[{11}\right]_U+g_{1}\left(U\left[{11}\right]_S+
2\left[U_{\cdot 1}S_{\cdot 1}\right]\right)+g_{2}U\left[{1}^{2}\right]_S,
\nonumber
\\
\left[1^{3}\right]_T
&=&
g_{0}^{3}\left[{1}^{3}\right]_U +3g_{0}^{2}g_{1}U\left[U_{\cdot 1}^{2}S_{\cdot 1}\right]
+3g_{0}g_{1}^{2}U^{2}\left[U_{\cdot 1}S_{\cdot 1}^{2}\right]+
g_{1}^{3}U^{3}\left[{1}^{3}\right]_S,
\nonumber
\\
\left[1,2,12\right]_T
&=&
g_{0}^{3}\left[{1},{2},{12}\right]_U +g_{0}^{2}g_{1}\left(U\left[U_{\cdot 1}U_{\cdot 2}
S_{\cdot 12}\right] +  2\left[{1}^{2}\right]_U\left[U_{\cdot 1}S_{\cdot 1}\right]+
2U\left[U_{\cdot 1}S_{\cdot 2}U_{\cdot 12}\right]\right)
\\
&&
+g_{0}^{2}g_{2}U\left[U_{\cdot 1} S_{\cdot 1}\right]^{2}
+  g_{0}g_{1}^{2}U\Bigg(2\left[U_{\cdot 1}S_{\cdot 1}\right]^{2}+
2\left[{1}^{2}\right]_U\left[{1}^{2}\right]_S+2U\left[U_{\cdot 1}
S_{\cdot 2}S_{\cdot 12}\right]
\nonumber
\\
&&
+U\left[S_{\cdot 1}S_{\cdot 2}U_{\cdot 12}\right]\Bigg)
+  2g_{0}g_{1}g_{2}U^{2}\left[U_{\cdot 1}S_{\cdot 1}\right]\left[{1}^{2}\right]_S
+g_{1}^{3}U^{2}\Bigg(U\left[{1},{2},{12}\right]_S
\nonumber
\\
&&
+ 2\left[U_{\cdot 1}S_{\cdot 1}\right]\left[{1}^{2}\right]_S\Bigg)+
g_{1}^{2}g_{2}U^{3}\left[{1}^{2}\right]_S^{2}.
\nonumber
\end{eqnarray*}
Similarly, the  cumulant coefficients $a_{22}$, $a_{43}$ needed for
third order inference are
given by (\ref{3.16}) in terms
of the bracket functions given in Appendix A.
The bracket functions needed for  (\ref{3.8}), (\ref{3.9})
for estimates of $T(F)$ of bias $O(n^{-3})$ are
\begin{eqnarray*}
\left[111\right]_T
&=&
g_{0}\left[{111}\right]_U+g_{1}\left(U\left[{111}\right]_S+
3\left[U_{\cdot 1}S_{\cdot 11}\right]+3\left[U_{\cdot 11} S_{\cdot 1}\right]\right)
\nonumber
\\
&&
+ 3g_{2}\left(U\left[{1},{11}\right]_S+
\left[U_{\cdot 1}S_{\cdot 1}^{2}\right]\right)+g_{3}U\left[{1}^{3}\right]_S,
\nonumber
\\
\left[1122\right]_T
&=&
g_{0}\left[{1122}\right]_U+g_{1}\Bigg(U\left[{1122}\right]_S+
4\left[U_{\cdot 1}S_{\cdot 122}\right]+
4\left[U_{\cdot 122}S_{\cdot 1}\right]
\nonumber
\\
&&
+ 2\left[{11}\right]_U\left[{11}\right]_S+4\left[U_{\cdot 12}S_{\cdot 12}\right]\Bigg)+
g_{2}\Bigg( 4U\left[{1},{122}\right]_S+U\left[{11}\right]_S^{2}
+2U\left[{12}^{2}\right]_S
\nonumber
\\
&&
+8\left[U_{\cdot 1}S_{\cdot 2}S_{\cdot 12}\right]+4\left[S_{\cdot 1}S_{\cdot 2}U_{\cdot 12}\right]+
4\left[U_{\cdot 1}S_{\cdot 1}\right]\left[{11}\right]_S+
2\left[{11}\right]_U\left[{1}^{2}\right]_S\Bigg)
\nonumber
\\
&&
+ 2g_{3}\left(U\left[{1}^{2}\right]_S\left[{11}\right]_S+
2U\left[{1},{2},{12}\right]_S+2\left[U_{\cdot 1}S_{\cdot 1}\right]
\left[{1}^{2}\right]_S\right)+g_{4}U\left[{1}^{2}\right]_S^{2}.
\nonumber
\end{eqnarray*}
Further terms are given in Appendix A.

If $g(s)=s^{r}$ then $g_{i}=(r)_{i}S^{r-i}$.
Putting $r=-1$ gives the derivatives of a quotient $(-1)_{i}=(-1)^{i}i!$.
\end{example}

\section{Applications to moments}
\setcounter{equation}{0}

Suppose $X\sim F$ on ${\mathbb{R}}$.
Set $\mu = \mathbb{E} [X]$, $\mu_{r}^{'}=\mathbb{E} [X^{r}]$ and let $\{ \mu_{r}, \kappa_{r}\}$ be
the central moments and cumulants of $F$.
Set $\mu (F)=\mu$ and so on.
Let $\widehat{F}$ be the empirical distribution of a random sample of size $n$ from $F$.

Many authors have studied problems of moments and cumulants: see, for example,  Stuart and Ord (1987).
Fisher's {\it k-statistic} $k_r$, the unbiased estimate of $\kappa_{r}$,
is given there in Section 12.9 for $r\leq 8$ in terms of
$\{ s_{i}=n\mu_{i}^{'}(\widehat{F})=\sum_{j=1}^{n}X_{j}^{i}\}$.
Fisher's expressions for unbiased estimates of the joint cumulants
of  $k$-statistics are given
there in Section 12.16.
Wishart's unbiased estimates of products of cumulants
are given there in Section 12.16 in  terms of symmetric functions, which can be converted to
$\{ s_{i}\}$ using Appendix Table 10.

Generally one
only wants approximations.
(Indeed without making parametric assumptions on $F$
only approximations are possible except for estimating polynomials in moments).
One problem with these ``traditional'' approaches is that
it is not
an easy task to separate out terms beyond the first in decreasing order of
importance in order to make such approximations.
As noted in Section 3 the present approach does not suffer from this disadvantage.

For $S(F)$ a polynomial in $F$ of degree $r$ (for example, $\mu_{r}^{'}$,
$\mu_{r}$ or $\kappa_{r}$), derivatives of order beyond $r$ vanish.

\begin{example}
Suppose $T(F)$ is a function of a univariate mean, say $T(F)=g(\mu (F))$.
Setting $g_{k}=g^{(k)}(\mu)$, Example 3.1 implies
\begin{eqnarray*}
&&
a_{21} = g_{1}^{2}\mu_{2},
\\
&&
a_{11} =  g_{2}\mu_{2}/2,
\\
&&
a_{32} = g_{1}^{3}\mu_{3}+3g_{1}^{2}g_{2}\mu_{2}^{2},
\\
&&
a_{22} = g_{1}g_{2}\mu_{3}+\left(g_{2}^{2}/2+g_{1}g_{3}\right)\mu_{2}^{2},
\\
&&
a_{43} =  g_{1}^{4} \left(\mu_{4}-3\mu_{2}^{2}\right)+12g_{1}^{3}g_{2}\mu_{3}\mu_{2}+
4\left(3g_{1}^{2}g_{2}^{2}+g_{1}^{3}g_{3}\right)\mu_{2}^{3}.
\end{eqnarray*}
For,
\begin{eqnarray}
T_{\cdot 1 \cdots p} = g_{p}h_{1} \cdots h_{p},
\nonumber
\end{eqnarray}
where $h_{i} = x_{i}-\mu$.
So,
\begin{eqnarray}
&&
\left[1^{k}\right] = g_{1}^{k}\mu_{k},
\nonumber
\\
&&
\left[1 \cdots 1\right] = g_{k}\mu_{k} \mbox{ if $1 \cdots 1$ contains $k$ 1's},
\nonumber
\\
&&
\left[1,2,12\right] = g_{1}^{2}g_{2}\mu_{2}^{2},
\nonumber
\\
&&
\left[1,11\right] = g_{1}g_{2}\mu_{3},
\nonumber
\\
&&
\left[12^2\right] = g_{2}^{2}\mu_{2}^{2},
\nonumber
\\
&&
\left[1,122\right] = g_{1}g_{3}\mu_{2}^{2},
\nonumber
\\
&&
\left[1,2^{2},12\right] = g_{1}^{3}g_{2}\mu_{3}\mu_{2},
\nonumber
\\
&&
\left[1,2,13,23\right] = g_{1}^{2}g_{2}^{2}\mu_{2}^{3},
\nonumber
\\
&&
\left[1,2,3,123\right] = g_{1}^{3}g_{3}\mu_{2}^{3}.
\nonumber
\end{eqnarray}
So, an estimate of $g(\mu (F))$ of bias
$O(n^{-4})$ is given by (\ref{3.7}) with $j=3$ in terms of
\begin{eqnarray*}
&&
S_{\cdot 1}(F) = -g_{2}\mu_{2}/2,
\\
&&
S_{\cdot 2}(F) = g_{3}\mu_{3}/3+g_{4}\mu_{2}^{2}/8,
\\
&&
S_{\cdot 3}(F) = -g_{4}\mu_{4}/4+3g_{4}\mu_{2}^{2}/8-g_{5}\mu_{3}\mu_{2}/6-g_{6}\mu_{2}^{3}/48.
\end{eqnarray*}
For example, an estimate of $\mu^{r}$ of bias
$O(n^{-4})$ is given by substituting $g_{i}=(r)_{i}\mu^{r-i}$.
If $\mu\geq 0$, $r$ need not be an integer.
However, regularity conditions generally breakdown if $r<0$ and $\dot{F}(0)\neq 0$.
\end{example}

Functions of non-central moments can be handled with similar ease.
We now present an important result which was stated without
proof in equation (4.1)
of Withers (1987), {\it  the derivatives of a central moment}.

\begin{theorem}
For $r$, $p$ in $\{ 1, 2, \ldots \}$, the $p$th derivative of $\mu_{r}(F)$ is
\begin{eqnarray}
\mu_{r \cdot 1 \cdots p} =
(-1)^{p}\left\{ (r)_{p}\mu_{r-p}-(r)_{p-1}\sum^{p}_{i=1}
\left( h_{i}^{r-p} - \mu_{r-p+1}h_{i}^{-1} \right)\right\}
\prod_{i=1}^{p}h_{i},
\nonumber
\end{eqnarray}
where $h_{i}=x_{i}-\mu$.
\end{theorem}

\noindent
{\bf Proof:}~~As in Example 3.1, $T(F)=\mu_{k}^{'}\mu^{j}$ has derivatives
\begin{eqnarray*}
T_{\cdot 1 \cdots p} =
\left( g_{p-1}\sum^{p}_{i = 1}U_{\cdot i}S_{\cdot i}^{-1}+g_{p}U\right)
S_{\cdot 1} \cdots S_{\cdot p},
\end{eqnarray*}
where
\begin{eqnarray*}
g_{p} = (j)_{p}\mu^{j-p},
\
U=\mu_k^{'},
\
U_{\cdot i}=x_{i}^{k}-\mu^{'}_{k},
\
S_{\cdot i}=x_{i}-\mu.
\end{eqnarray*}
But $\mu_{r}=\sum_{k = 0}^r {r \choose k}(-1)^{r-k}\mu_{k}^{'}\mu^{r-k}$.
So,
\begin{eqnarray*}
\mu_{r \cdot 1 \cdots p}
&=&
\sum_{k = 0}^r {r \choose k}(-1)^{r-k}\Bigg\{ (r-k)_{p-1}\mu^{r-k-p+1}
\sum_{i=1}^{p} \left(x_{i}^{k}-\mu_{k}^{'}\right) h_{i}^{-1}
\\
&&
\qquad \qquad \qquad
+(r-k)_{p}\mu^{r-k-p}\mu_{k}^{'}\Bigg\} \prod_{j = 1}^{p}h_{j}.
\end{eqnarray*}
Now simplify.
\
$\Box$

Some particular cases of the theorem are given by the following corollaries.

\begin{corollary}
We have
\begin{eqnarray*}
&&
\mu_{r \cdot 1} = h_1^r-\mu_r- r h_1\mu_{r-1},
\\
&&
\mu_{r \cdot 12} = -r\sum^2 \left(h_1^{r-1}-\mu_{r-1}\right)
h_2 +(r)_2h_1h_2 \mu_{r-2},
\\
&&
\mu_{r \cdot 123} = (r)_2\sum^3 \left(h_1^{r-2}-\mu_{r-2}\right)
h_2h_3 -(r)_3h_1h_2h_3 \mu_{r-3},
\\
&&
\mu_{r \cdot 12\cdots r-1} = (-1)^{r-1} \left(r!/2\right)
\sum^{r-1} \left(h_1^2-\mu_2\right) h_2\cdots h_{r-1},
\\
&&
\mu_{r \cdot 12\cdots r} = (-1)^r (r-1) r! h_1\cdots h_r,
\nonumber
\\
&&
\mu_{r \cdot 1^p} = (-1)^{p} \left\{ (r)_p\mu_{r-p}h_1^p  - p(r)_{p-1}
h_1^r-\mu_{r-p+1}h_1^{p-1} \right\}.
\end{eqnarray*}
\end{corollary}

\begin{corollary}
We have
\begin{eqnarray*}
&&
\mu_{2 \cdot 1} = h_{1}^{2}-\mu_{2},
\\
&&
\mu_{3 \cdot 1} = h_1^3-\mu_3-3 h_1\mu_2,
\\
&&
\mu_{4 \cdot 1} = h_1^4-\mu_4-4 h_1\mu_3,
\\
&&
\mu_{5 \cdot 1} = h_1^5-\mu_5-5 h_1\mu_4,
\\
&&
\mu_{2 \cdot 12} = -2h_1h_2,
\\
&&
\mu_{3 \cdot 12} = -3\sum^{2}\left(h_{1}^{2}-\mu_{2}\right)h_{2},
\\
&&
\mu_{4 \cdot 12} = 12h_1h_2\mu_2-4\sum^{2}\left(h_{1}^{3}-\mu_{3}\right)h_{2},
\\
&&
\mu_{5 \cdot 12} = 20h_1h_2\mu_3-5\sum^{2}\left(h_{1}^{4}-\mu_{4}\right)h_{2},
\\
&&
\mu_{3 \cdot 123} = 12 h_1h_2 h_3,
\\
&&
\mu_{4 \cdot 123} = 12\sum^{3} \left(h_1^2-\mu_2\right) h_2h_3,
\\
&&
\mu_{5 \cdot 123} = 20\sum^{3} \left(h_1^3-\mu_3\right) h_2h_3-60h_1h_2h_3,
\\
&&
\mu_{4 \cdot 1234} = 72 h_1h_2 h_3h_4,
\\
&&
\mu_{5 \cdot 1234} = 60\sum^{4} \left(h_1^2-\mu_2\right) h_2h_3h_4,
\\
&&
\mu_{5 \cdot 12345} = -480 h_1h_2 h_3h_4h_5.
\end{eqnarray*}
\end{corollary}

So, for example, for $T(F)=\mu_r$,
\begin{eqnarray*}
\left[1^{3}\right]_T
&=&
\mu_{3r}-3r\mu_{2r+1}\mu_{r-1}-3\mu_{2r}\mu_{r}+3r^{2}
\mu_{r+2}\mu_{r-1}^{2}+6r\mu_{r+1}\mu_{r}\mu_{r-1}
\nonumber
\\
&&
+2\mu_{r}^{3}-3r^{2}\mu_{r}\mu_{r-1}^{2}\mu_{2}-r^{3}\mu_{r-1}^{3} \mu_{3},
\nonumber
\\
\left[1,2,12\right]_T
&=&
-\mu_{2r-1}\left( 2r\mu_{r+1}+r^{2}\mu_{r-1}\mu_{2}\right)+
(r)_{2}\mu_{r+1}^{2}\mu_{r-2}
\nonumber
\\
&&
-r^{2}\mu_{r+1}\mu_{r}\mu_{r-1}+2r^{2}(r-1)\mu_{r+1}\mu_{r-1}\mu_{r-2}\mu_{2}
\nonumber
\\
&&
-\left(2r^{3}-r^{2}\right)\mu_{r}\mu_{r-1}^{2}\mu_{2}+r^{3}(r-1)\mu_{r-1}^{2}\mu_{r-2} \mu_{2}^{2},
\nonumber
\end{eqnarray*}
giving
\begin{eqnarray}
a_{21}
&=&
\left[1^{2}\right]_T=r^{2}\mu_{r-1}^{2}\mu_{2}-2r\mu_{r-1}\mu_{r+1}+\mu_{2r}-\mu_{r}^{2},
\nonumber
\\
a_{11}
&=&
[11]_T/2=(r)_{2}\mu_{r-2}\mu_{2}/2-r\mu_{r},
\nonumber
\\
a_{32}
&=&
\mu_{3r}-3r\mu_{2r+1}\mu_{r-1}-3\mu_{2r}\mu_{r}-3
\mu_{2r-1}\left(2r\mu_{r+1}+r^{2}\mu_{r-1}\mu_{2}\right)
\nonumber
\\
&&
+3r^{2}\mu_{r+2}\mu_{r-1}^{2}+3(r)_{2}\mu_{r+1}^{2}\mu_{r-2}-3r(r-2)
\mu_{r+1}\mu_{r}\mu_{r-1}
\nonumber
\\
&&
+6r^{2}(r-1)\mu_{r+1}\mu_{r-1}\mu_{r-2}\mu_{2}+2\mu_{r}^{3}-3r^{2}\mu_{r}
\mu_{r-1}^{2}\mu_{2}-r^{3}\mu_{r-1}^{3}\mu_{3}.
\nonumber
\end{eqnarray}
Similarly, estimates of $\mu_{r}$ for {\it general} $r$
of bias $O(n^{-3})$ are given by (\ref{3.7}) in terms of
\begin{eqnarray*}
&&
[111]_T = -(r)_{3}\mu_{r-3}\mu_{3}+3(r)_{2} \left( \mu_{r}-\mu_{r-2} \mu_{2}\right),
\nonumber
\\
&&
[1122]_T = (r)_{4}\mu_{r-4}\mu_{2}^{2}.
\nonumber
\end{eqnarray*}
For $r=2$ this gives
\begin{eqnarray*}
a_{21} = \mu_4-\mu_2^2,
\
a_{11}=-\mu_2,
\
a_{32} = \mu_6-3\mu_4\mu_2+2\mu_2^3,
\
[111]_T  = [1122]_T=0,
\end{eqnarray*}
and for $r=3$ this gives
\begin{eqnarray*}
&&
a_{21} = \mu_6-4\mu_4\mu_2-\mu_3^2+9\mu_2^3,
\
a_{11}=-3\mu_3,
\\
&&
a_{32}  = \mu_9-9\mu_7\mu_2-3\mu_6\mu_3-18\mu_5\mu_4-9\mu_4\mu_3\mu_2+2\mu_3^3,
\\
&&
[111]_T  = 12\mu_3,
\
[1122]_T=0.
\end{eqnarray*}

\begin{example}
Suppose that $r$ is an odd integer and $F$ is symmetric.
So, odd cumulants
of $\mu_r(\widehat{F})$ are  zero so that
$a_{11} =  a_{32} =0$ and the Edgeworth-Cornish-Fisher expansions are in
powers of $n^{-1}$, not just $n^{-1/2}$.
Taking  $r=3$ gives for $T(F)=\mu_3$,
\begin{eqnarray*}
&&
\left[12^2\right]_T = 2 \mu_2\left( \mu_4- \mu_2^2\right),
\\
&&
\left[1,11\right]_T = -6\left( \mu_6-4 \mu_4 \mu_2+3 \mu_2^3\right),
\\
&&
\left[1,122\right]_T = 12  \mu_2\left( \mu_4-3 \mu_2^2\right),
\\
&&
\left[1^4\right]_T =  \mu_{12}-12 \mu_{10} \mu_2+5 \mu_8 \mu_2^2-108 \mu_6 \mu_2^3+81 \mu_4 \mu_2^4,
\\
&&
\left[1,2^2,12\right]_T = -3 \mu_4 \left( \mu_8-4 \mu_6 \mu_2+6 \mu_4 \mu_2^2-3 \mu_2^4\right),
\\
&&
\left[1,2,13,23\right]_T = 9\left( \mu_4- \mu_2^2\right)\left( \mu_4-3 \mu_2^2\right)^2.
\end{eqnarray*}
So,
\begin{eqnarray*}
&&
a_{21} =  \mu_6-6 \mu_4 \mu_2+9 \mu_2^3,
\\
&&
a_{22} = -6 \left( \mu_6-7 \mu_4 \mu_2+10 \mu_2^3\right),
\\
&&
a_{43} =  \mu_{12}-12 \mu_{10} \mu_2 - \mu_8 \left(72 \mu_4-5 \mu_2^2\right) -
3 \mu_6\left( \mu_6-108 \mu_4 \mu_2+54 \mu_2^3\right)
\\
&&
\qquad + 3 \mu_4 \left(52 \mu_4^2-576 \mu_4 \mu_2^2+1179 \mu_2^4 \right) - 2511 \mu_2^6.
\end{eqnarray*}
For $F$ normal this gives $a_{21} = 6\mu_2^3$, $a_{22} = -24\mu_2^3$, $a_{43} = -11625\mu_2^6$.
\end{example}

\begin{example}
This example is about standardized central moments.
Suppose $T(F)=\nu_r$, where $\nu_r=\mu_{r}/\mu_{2}^{r/2}$.
Then the $[\cdot ]_T$ needed for third order inference and bias reduction,
are given by Example 3.4 with $S=\mu_{2}$ and $U=\mu_{r}$ and
\begin{eqnarray*}
g_{j}=(-r/2)_{j}\mu_{2}^{-r/2-j}
=r(r+2)(r+4) \cdots (r+2j-2)\left(-2\mu_{2}\right)^{-j} \mu_{2}^{-r/2}
\end{eqnarray*}
in terms of $[{1}^{2}]_U$, $[{11}]_U$, $[{1}^{3}]_U$, $\cdots$ and
$[{1}^{2}]_S$, $[{11}]_S$, $[{1}^{3}]_S$, $\cdots $ given by Example 4.2
and the bracket functions
\begin{eqnarray*}
&&
\left[U_{\cdot 1}S_{\cdot 1}\right] = \mu_{r+2}-2\mu_{r}\mu_{2}-r\mu_{r-1}   \mu_{3},
\nonumber
\\
&&
\left[U_{\cdot 1}^{2}S_{\cdot 1}\right] = \mu_{2r+2}-\mu_{2r}\mu_{2}-2r\mu_{r-1}
\left(\mu_{r+3}-\mu_{r+1}\mu_{2}-\mu_{r}\mu_{3}\right)-2\mu_{r+2}\mu_{r}
\nonumber
\\
&&
\qquad
+2\mu_{r}^{2}\mu_{2}+r^{2}\mu_{r-1}^{2}\left(\mu_{4}-\mu_{2}^{2}\right),
\nonumber
\\
&&
\left[U_{\cdot 1}S_{\cdot 1}^{2}\right] = \mu_{r+4}-2\mu_{r+2}\mu_{2}+\mu_{r}\left(-\mu_{4}
+2\mu_{2}^{2}\right)-r\mu_{r-1}\left(\mu_{5}+2\mu_{3}\mu_{2}\right),
\nonumber
\\
&&
\left[U_{\cdot 1}U_{\cdot 2}S_{\cdot 12}\right] = -2\left(\mu_{r+1}-r\mu_{r-1}\mu_{2}\right)^{2},
\nonumber
\\
&&
\left[U_{\cdot 1}S_{\cdot 2}U_{\cdot 12}\right] = -r\mu_{2r-1}\mu_{3}-r\mu_{r+1}^{2}+
\left(r^{2}+r\right)\mu_{r+1}\mu_{r-1}\mu_{2}+\left(r\right)_{2}\mu_{r+1}\mu_{r-2}\mu_{3}
\nonumber
\\
&&
\qquad
+\left(r^{2}+r\right)\mu_{r}\mu_{r-1}\mu_{3}-r^{2}\mu_{r-1}^{2}\mu_{2}^{2}-r^{2}\left(r-1\right)
\mu_{r-1}\mu_{r-2}\mu_{3}\mu_{2},
\nonumber
\\
&&
\left[U_{\cdot 1}S_{\cdot 2}S_{\cdot 12}\right] =
2\left(-\mu_{r+1}\mu_{3}+r\mu_{r-1}\mu_{3}  \mu_{2}\right),
\nonumber
\\
&&
\left[S_{\cdot 1}S_{\cdot 2}U_{\cdot 12}\right] = -2r\mu_{r+1}\mu_{3}+4r\mu_{r-1}\mu_{3}\mu_{2}
+\left(r\right)_{2}\mu_{r-2}\mu_{3}^{2},
\nonumber
\\
&&
\left[U_{\cdot 1}S_{\cdot 11}\right] = 2\left(-\mu_{r+2}+\mu_{r}\mu_{2}+r\mu_{r-1}\mu_{3}\right),
\nonumber
\\
&&
\left[U_{\cdot 11}S_{\cdot 1}\right] = 2r\left(-\mu_{r+2}+\mu_{r}\mu_{2}+\mu_{r-1}\mu_{3}\right)+
\left(r\right)_{2}\mu_{r-2}\left(\mu_{4}-\mu_{2}^{2}\right),
\nonumber\\
&&
\left[U_{\cdot 1}S_{\cdot 122}\right] = 0,
\nonumber
\\
&&
\left[U_{\cdot 122}S_{\cdot 1}\right] = \left(r\right)_{2}\left(\mu_{r}\mu_{2}+2\mu_{r-1}\mu_{3}-\mu_{r-2}
\mu_{2}^{2}-\left(r-2\right)\mu_{r-3}\mu_{3}\mu_{2}\right),
\nonumber
\\
&&
\left[U_{\cdot 12}S_{\cdot 12}\right] = 4r\mu_{r}\mu_{2}-2\left(r\right)_{2}\mu_{r-2} \mu_{2}^{2}.
\nonumber
\end{eqnarray*}
For example, suppose that $r=3$ and $F$ is symmetric.
Then $a_{ri}=0$ for $r$ odd and
\begin{eqnarray*}
a_{21}
&=&
\nu_6-6\nu_4+9,
\\
a_{22}
&=&
-3\left(\nu_8-5\nu_6+7\nu_4-3\right)+12 \nu_6\left(2 \nu_4-1\right)/4
\\
&&
+2\nu_4\left(107\nu_4-489\right)/4+9\left(4\nu_4-11\right),
\\
a_{43}
&=&
\nu_{12}-12\nu_{10}+54\nu_8-108\nu_6+81\nu_4-3a_{21}^2
\\
&&
-18\left(\nu_8-4\nu_6+3\nu_4-3\right)\left(\nu_6-4\nu_4+9\right)
\\
&&
+27\left(\nu_4-1\right) \left[a_{21}^2 +4a_{21}\left(\nu_4-3\right)+
4\left(\nu_6\left(\nu_4-1\right)+9\right)\right]
\\
&&
+12\left(\nu_4-3\right)^2\left(3\nu_6-14\nu_4+15\right).
\end{eqnarray*}
For $F$ normal this is in agreement with Fisher (1931) who gave the result
\begin{eqnarray*}
\mu\left(-r,a_3,a_4,\ldots\right) =
\mu \left(a_3,a_4,\ldots\right)(n-1)^r/\left\{(n-1)(n+1)\cdots (n+2r-3)\mu_2^r\right\}
\end{eqnarray*}
for $\mu(a_2,a_3,a_4,\ldots) = \mathbb{E} [k_2^{a_2} k_3^{a_3}\cdots]$.
See Agostino and Pearson (1973) for a simulation approach.

Figure 4.1 compares the bias  reduced estimator of $\nu_3$ versus the usual one by means of simulation.
The biases of the estimators are computed by simulating ten thousand replications of samples of size $n$ from the
following distributions: standard normal, Student's $t$ with two degrees of freedom,
Student's $t$ with five degrees of freedom, Student's $t$ with ten degrees of freedom,
standard logistic, standard Laplace.
As expected, the bias reduced estimators give substantially smaller biases for each $n$ and
for each of the six distributions.
The biases appear largest for the Student's $t$ distribution with two degrees of freedom.
The biases appear smallest for the normal distribution,
the Student's $t$ distribution with ten degrees of freedom, and the logistic distribution.
\end{example}

As noted $k_r$ is the unbiased estimate of $\kappa_r$ so
$k_2=\mu_2(\widehat{F})n/(n-1)$, and $k_3=\mu_3(\widehat{F})n^2/(n-1)(n-2)$.

\begin{example}
Suppose $T(F)=\mu_{r}/\mu^{r}$, where $\mu \neq 0$.
Then the $[\cdot ]_T$ needed for third order inference and bias reduction
are given by Example 3.4 with $g_{j}=(-r)_{j}\mu^{-r-j}$, $S=\mu$, $U=\mu_{r}$,
$[{1}^{2}]_U$, $[{11}]_U$, $[{1}^{3}]_U$, $\ldots $ given by Example 4.2,
$[{1}^{i}]_S=\mu_{i}$,  the other non-zero leading terms needed for Example 3.3 being
\begin{eqnarray*}
&&
\left[U_{\cdot 1}S_{\cdot 1}\right] = \mu_{r+1}-r\mu_{r-1}\mu_{2},
\nonumber
\\
&&
\left[U_{\cdot 1}^{2}S_{\cdot 1}\right] = \mu_{2r+1}-2r\mu_{r+2}\mu_{r-1}-2\mu_{r+1}
\mu_{r}+2r\mu_{r}\mu_{r-1}\mu_{2}+r^{2}\mu_{r-1}^{2}\mu_{3},
\nonumber
\\
&&
\left[U_{\cdot 1}S_{\cdot 1}^{2}\right] = \mu_{r+2}-\mu_{r}\mu_{2}-r\mu_{r-1}\mu_{3},
\nonumber
\\
&&
\left[U_{\cdot 1}S_{\cdot 2}U_{\cdot 12}\right] = -r\mu_{2r-1}\mu_{2}-r\mu_{r+1}\mu_{r}+r(2r+1)
\mu_{r}\mu_{r-1}\mu_{2}
\nonumber
\\
&&
\qquad
+ (r)_{2}\mu_{r-2}\left(\mu_{r+1}\mu_{2}-r\mu_{r-1}\mu_{2}^{2}\right),
\nonumber
\\
&&
\left[S_{\cdot 1}S_{\cdot 2}U_{\cdot 12}\right] = -2r\mu_{r}\mu_{2}+(r)_{2}\mu_{r-2} \mu_{2}^{2},
\nonumber
\\
&&
\left[U_{\cdot 11}S_{\cdot 1}\right] = -2r\left(\mu_{r+1}-\mu_{r-1}\mu_{2}\right)+(r)_{2}\mu_{r-2} \mu_{3},
\nonumber
\\
&&
\left[U_{\cdot 122}S_{\cdot 1}\right] = 3(r)_{2}\mu_{r-1}\mu_{2}-(r)_{3}\mu_{r-3}  \mu_{2}^{2}.
\nonumber
\end{eqnarray*}
For example, the asymptotic variance of $n^{1/2}(T(\widehat{F})-T(F))$ is
\begin{eqnarray*}
&&
\mu^{-2r}\left(r^{2}\mu_{r-1}^{2}\mu_{2}-2r\mu_{r-1}\mu_{r+1}
+\mu_{2r}-\mu_{r}^{2}\right)
\\
&&
- 2r\mu^{-2r-1} \left(\mu_{r+1}-r\mu_{r-1}\mu_{2}\right)
+r^{2}\mu^{-2r-2}\mu_{r}^{2}\left(\mu_{4}-\mu_{2}^{2}\right).
\end{eqnarray*}
For $r=2$, this reduces to
$T(F)^{2}(\mu_{4}\mu_{2}^{-2}-1-4\mu_{3}\mu_{2}^{-1}\mu^{-1}+4\mu_{2}\mu^{-2})$.
\end{example}

\begin{example}
This example is about the coefficient of variation.
Suppose $T(F)=\mu_{2}^{1/2}/\mu$.
Then the $[\cdot ]_T$ needed for third order inference and bias reduction
are given by Example 3.4 with $g_{j}=(1/2)_{j}\mu_{2}^{1/2-j}$, $S=\mu_{2}$, $U=\mu^{-1}$.
By Example 4.1, $U_{\cdot 1 \cdots p}=(-1)_{p} \mu^{-1-p}h_{1} \cdots h_{p}$
so the terms needed in Example 3.4 are
\begin{eqnarray*}
&&
\left[{1}^{2}\right]_U = \mu^{-4}\mu_{2},
\
\left[U_{\cdot 1}S_{\cdot 1}\right] = -\mu^{-2}\mu_{3},
\
\left[{1}^{2}\right]_S=\mu_{4}-\mu_{2}^{2},
\nonumber
\\
&&
\left[{11}\right]_U = 2\mu^{-3}\mu_{2},
\
\left[{11}\right]_S = -2\mu_{2},
\
\left[{1}^{3}\right]_U=  -\mu^{-6}\mu_{3},
\nonumber
\\
&&
\left[U_{\cdot 1}^{2}S_{\cdot 1}\right] = \mu^{-4}\left(\mu_{4}-\mu_{2}^{2}\right),
\
\left[U_{\cdot 1}S_{\cdot 1}^{2}\right] = -\mu^{-2}\left(\mu_{5}-2\mu_{3}\mu_{2}\right),
\nonumber
\\
&&
\left[{1}^{3}\right]_S = \mu_{6}-3\mu_{4}\mu_{2}+2\mu_{2}^{3},
\
\left[{1},{2},{12}\right]_U = 2\mu^{-7}\mu_{2}^{2},
\nonumber
\\
&&
\left[U_{\cdot 1}U_{\cdot 2}S_{\cdot 12}\right] = -2\mu^{-4}\mu_{2}^{2},
\
\left[U_{\cdot 1}S_{\cdot 2}U_{\cdot 12}\right] =-2\mu^{-5}\mu_{3}\mu_{2},
\nonumber
\\
&&
\left[U_{\cdot 1}S_{\cdot 2}S_{\cdot 12}\right] = 2\mu^{-2}\mu_{3}\mu_{2},
\nonumber
\\
&&
\left[S_{\cdot 1}S_{\cdot 2}U_{\cdot 12}\right] = 2\mu^{-3}\mu_{3}^{2},
\
\left[{1},{2},{12}\right]_S= -2\mu_{3}^{2},
\nonumber
\\
&&
\left[{111}\right]_U = -6\mu^{-4}\mu_{3},
\
\left[{111}\right]_S = 0,
\
\left[U_{\cdot 1}S_{\cdot 11}\right]=2\mu^{-2}  \mu_{3},
\nonumber
\\
&&
\left[U_{\cdot 11}S_{\cdot 1}\right] = 2\mu^{-3}\left(\mu_{4}-\mu_{2}^{2}\right),
\
\left[S_{\cdot 1}S_{\cdot 11}\right]= -2\left(\mu_{4}-\mu_{2}^{2}\right),
\nonumber
\\
&&
\left[{1122}\right]_U = 24\mu^{-5}\mu_{2}^{2},
\
\left[{1122}\right]_S = 0,
\
\left[U_{\cdot 1} S_{\cdot 122}\right]=0,
\nonumber
\\
&&
\left[U_{\cdot 122}S_{\cdot 1}\right] = -6\mu^{-4}\mu_{2}^{2},
\
\left[U_{\cdot 12}S_{\cdot 12}\right]= -4\mu^{-3}\mu_{2}^{2},
\nonumber
\\
&&
\left[{1},{122}\right]_S = 0,
\
\left[{11}^{2}\right]_S = 4\mu_{4},
\
\left[{12}^{2}\right]_S=  4\mu_{2}^{2},
\nonumber
\\
&&
\left[U_{\cdot 1}S_{\cdot 1}S_{\cdot 12}\right] = 2\mu^{-2}\mu_{3}\mu_{2}.
\nonumber
\end{eqnarray*}
For example,
\begin{eqnarray*}
a_{21} = \left[ 1^{2}\right] = T(F)^{2}
\left( \mu_{2}\mu^{-2}-\mu_{3}\mu^{-1}\mu_{2}^{-1}+ \mu_{4}\mu_{2}^{-2}4^{-1}-4^{-1}\right)
\end{eqnarray*}
as given by Section 10.6 of  Stuart and Ord (1987).
Also
\begin{eqnarray*}
&&
a_{11} = [11]/2=T(F)\left(-3/8-\mu_{2}^{-2}\mu_{4}/8-\mu^{-1}\mu_{2}^{-1} \mu_{3}/2+\mu^{-2}\mu_{2}\right),
\\
&&
\left[1^{3}\right] = T(F)^{3}\sum_{i = 0}^{3}\mu^{-i}A_{i},
\\
&&
[1,2,12] = T(F)^{3}\sum_{i = 0}^{4}\mu^{-i}B_{i},
\\
&&
a_{32} = T(F)^{3}\sum_{i = 0}^{4}\mu^{-i}C_{i},
\end{eqnarray*}
where
\begin{eqnarray*}
&&
A_{0}=1/4 -3\mu_{2}^{-2}\mu_{4}/8+\mu_{2}^{-3}\mu_{6}/8,
\nonumber
\\
&&
A_{1}=3\left(2\mu_{2}^{-1}\mu_{3}-\mu_{2}^{-2}\mu_{5}\right)/4,
\
A_{2}=3\left(-\mu_{2}  +\mu_{2}^{-1}\mu_{4}\right)/2,
\
A_{3}=-\mu_{3},
\\
&&
B_{0}=-\left(1-\mu_{2}^{-2}\mu_{4}\right)^{2}/16-\mu_{2}^{-2}\mu_{3}^{2}/4,
\nonumber
\\
&&
B_{1}=\mu_{2}^{-1}\mu_{3}/2,
\
B_{2}=-3\mu_{2}/2+\mu_{2}^{-1}\mu_{4}/2+3\mu_{2}^{-2}\mu_{3}^{2}/4,
\
B_{3}=-3\mu_{3},
\nonumber
\\
&&
B_{4}=2\mu_{2}^{2},
\nonumber
\\
&&
C_{0}=1/16-3\mu_{2}^{-2}\mu_{3}^{2}/4+\mu_{2}^{-3}\mu_{6}/8-3\mu_{2}^{-4}\mu_{4}^{2}/16,
\nonumber
\\
&&
C_{1}=3\left(\mu_{2}^{-1}\mu_{3}-\mu_{2}^{-2}\mu_{5}/4\right),
\
C_{2}=3\left(-\mu_{2}+\mu_{2}^{-1}\mu_{4}+3\mu_{2}^{-2}\mu_{3}^{2}/4\right),
\nonumber
\\
&&
C_{3}=-10\mu_{3},
\
C_{4}=6\mu_{2}^{2}.
\nonumber
\end{eqnarray*}
\end{example}

\section*{Appendix A}
\renewcommand{\theequation}{$\mbox{A.\arabic{equation}}$}
\setcounter{equation}{0}

Continuing Example 3.4, the terms needed for  (\ref{3.16})-(\ref{3.17}) are:
\begin{eqnarray*}
\left[1,11\right]_T
&=&
g_{0}^{2}\left[{1},{11}\right]_U
+g_{0}g_{1}\Bigg(U\left[S_{\cdot 1}U_{\cdot 11}\right]+U\left[U_{\cdot 1} S_{\cdot 11}\right]
\nonumber
\\
&&
+2\left[U_{\cdot 1}^{2}S_{\cdot 1}\right]\Bigg)
+g_{0}g_{2}U\left[U_{\cdot 1}S_{\cdot 1}^{2}\right]+g_{1}g_{2}U^{2}\left[{1}^{3}\right]_S,
\nonumber
\\
\left[12^{2}\right]_T
&=&
g_{0}^{2}\left[{12}^{2}\right]_U+2g_{0}g_{1}\Bigg(U\left[U_{\cdot 12}S_{\cdot 12}\right]+2
\left[U_{\cdot 1}S_{\cdot 2}U_{\cdot 12}\right]\Bigg)
\nonumber
\\
&&
+g_{1}^{2}\Bigg(U^{2}\left[{12}^{2}\right]_S+4U\left[U_{\cdot 1}S_{\cdot 2}S_{\cdot 12}\right]+
2\left[{1}^{2}\right]_U\left[{1}^{2}\right]_S
+2\left[U_{\cdot 1}S_{\cdot 1}\right]^{2}\Bigg)
\nonumber
\\
&&
+2g_{0}g_{2}U\left[S_{\cdot 1}S_{\cdot 2}U_{\cdot 12}\right]+
2g_{1}g_{2}U\Bigg(U\left[{1},{2},{12}\right]_S+2\left[U_{\cdot 1}
S_{\cdot 1}\right]\left[{1}^{2}\right]_S\Bigg)
\nonumber
\\
&&
+g_{2}^{2}U^{2}\left[{1}^{2}\right]_S^{2}\Bigg),
\nonumber
\\
\left[1,122\right]_T
&=&
g_{0}^{2}\left[{1},{122}\right]_U+g_{0}g_{1}\Bigg(U\left[U_{\cdot 1}S_{\cdot 122}\right]+
\left[{1}^{2}\right]_U\left[{11}\right]_S+2\left[U_{\cdot 1}U_{\cdot 2}S_{\cdot 12}\right]
\nonumber
\\
&&
+2\left[U_{\cdot 1}S_{\cdot 2}U_{\cdot 12}\right]+
\left[U_{\cdot 1}S_{\cdot 1}\right]\left[{11}\right]_U+U\left[S_{\cdot 1}U_{\cdot 122}\right]\Bigg)
\nonumber
\\
&&
+g_{0}g_{2}\Bigg(U\left[U_{\cdot 1}S_{\cdot 1}\right]\left[{11}\right]_S+
2U\left[U_{\cdot 1}S_{\cdot 2}S_{\cdot 12}\right]
+\left[{1}^{2}\right]_U\left[{1}^{2}\right]_S+2\left[U_{\cdot 1}S_{\cdot 1}\right]^{2}\Bigg)
\nonumber
\\
&&
+g_{0}g_{3}U\left[U_{\cdot 1}S_{\cdot 1}\right]\left[{1}^{2}\right]_S+
g_{1}^{2}U\Bigg(U\left[{1},{122}\right]_S+\left[S_{\cdot 1}U_{\cdot 1}\right]\left[{11}\right]_S
\nonumber
\\
&&
+2\left[U_{\cdot 1}S_{\cdot 2}S_{\cdot 12}\right]+\left[{1}^{2}\right]_S\left[{11}\right]_U+
2\left[S_{\cdot 1}S_{\cdot 2}U_{\cdot 12}\right]\Bigg)+
g_{1}g_{2}
U\Bigg(U\left[{1}^{2}\right]_S\left[{11}\right]_S
\nonumber
\\
&&
+2U\left[{1},{2},{12}\right]_S+\left[S_{\cdot 1}U_{\cdot 1}\right]\left[{1}^{2}\right]_S+
2\left[{1}^{2}\right]_S\left[U_{\cdot 1}S_{\cdot 1}\right]\Bigg)
+2g_{1}g_{3}U^{2}\left[{1}^{2}\right]_S^{2}\Bigg),
\nonumber
\\
\left[1^{4}\right]_T
&=&
g_{0}^{4}\left[{1}^{4}\right]_U+4g_{0}^{3}g_{1}U\left[U_{\cdot 1}^{3}S_{\cdot 1}\right]+
6g_{0}^{2}g_{1}^{2}U^{2}\left[U_{\cdot 1}^{2}S_{\cdot 1}^{2}\right]
\nonumber
\\
&&
+4g_{0}g_{1}^{3}U^{3}\left[U_{\cdot 1}S_{\cdot 1}^{3}\right]+g_{1}^{4}U^{4}\left[{1}^{4}\right]_S,
\nonumber
\end{eqnarray*}
\begin{eqnarray*}
\left[1,2^{2},12\right]_T
&=&
g_{0}^{4}\left[{1},{2}^{2},{12}\right]_U+g_{0}^{3}g_{1}
\Bigg(U\left[U_{\cdot 1}U_{\cdot 2}^{2}S_{\cdot 12}\right]+
\left[{1}^{2}\right]_U\left[U_{\cdot 1}^{2}S_{\cdot 1}\right]
\nonumber
\\
&&
+\left[{1}^{3}\right]_U\left[U_{\cdot 1}S_{\cdot 1}\right]+
2U\left[U_{\cdot 1}U_{\cdot 2}S_{\cdot 2}U_{\cdot 12}\right]+U\left[U_{\cdot 1}^{2}S_{\cdot 2}
U_{\cdot 12}\right]\Bigg)
\nonumber
\\
&&
+g_{0}^{3}g_{2}U\left[U_{\cdot 1}S_{\cdot 1}\right]\left[U_{\cdot 1}^{2}S_{\cdot 1}\right]
\nonumber
\\
&&
+g_{0}^{2}g_{1}^{2}U\Bigg(2U\left[U_{\cdot 1}U_{\cdot 2}S_{\cdot 2}S_{\cdot 12}\right]+
2\left[{1}^{2}\right]_U\left[U_{\cdot 1}S_{\cdot 1}^{2}\right]
+2\left[U_{\cdot 1}S_{\cdot 1}\right]\left[U_{\cdot 1}^{2}S_{\cdot 1}\right]
\nonumber
\\
&&
+U\left[U_{\cdot 1}S_{\cdot 2}^{2}U_{\cdot 12}\right]+U\left[U_{\cdot 1}^{2}S_{\cdot 2}S_{\cdot 12}\right]
+\left[U_{\cdot 1}S_{\cdot 1}\right]\left[U_{\cdot 1}^{2}
S_{\cdot 1}\right]+\left[{1}^{3}\right]_U\left[{1}^{2}\right]_S\Bigg)
\nonumber
\\
&&
+g_{0}^{2}g_{1}g_{2}U^{2}\Bigg(2\left[U_{\cdot 1}S_{\cdot 1}\right]\left[U_{\cdot 1}S_{\cdot 1}^{2}\right]+
\left[U_{\cdot 1}^{2}S_{\cdot 1}\right]
\left[{1}^{2}\right]_S\Bigg)
\nonumber\\
&&
+g_{0}g_{1}^{3}U^{2}\Bigg(U\left[U_{\cdot 1}S_{\cdot 2}^{2}U_{\cdot 12}\right]+
\left[{1}^{2}\right]_U\left[{1}^{3}\right]_S
+\left[U_{\cdot 1}S_{\cdot 1}\right]\left[U_{\cdot 1}S_{\cdot 1}^{2}\right]+
2U\left[S_{\cdot 1}S_{\cdot 2}^{2}U_{\cdot 12}\right]\Bigg)
\nonumber
\\
&&
+g_{0}g_{1}^{2}g_{2}U^{3}\left[U_{\cdot 1}S_{\cdot 1}\right]\left[{1}^{3}\right]_S
\nonumber
\\
&&
+2g_{1}^{4}U^{3}\Bigg(U\left[{1},{2}^{2},{12}\right]_S+
\left[U_{\cdot 1}S_{\cdot 1}\right]\left[{1}^{3}\right]_S+\left[U_{\cdot 1}
S_{\cdot 1}^{2}\right]\left[{1}^{2}\right]_S\Bigg)
\nonumber
\\
&&
+ 2g_{1}^{3}g_{2}U^{4}\left[{1}^{2}\right]_S\left[{1}^{3}\right]_S\Bigg),
\nonumber
\end{eqnarray*}
\begin{eqnarray*}
\left[1,2,13,23\right]_T
&=&
g_{0}^{4}\left[{1},{2},{13},{23}\right]_U
\nonumber
\\
&&
+g_{0}^{3}g_{1}\Bigg(2U\left[U_{\cdot 1}S_{\cdot 2}U_{\cdot 13}U_{\cdot 23}\right]+
2U\left[U_{\cdot 1}U_{\cdot 2}U_{\cdot 13}S_{\cdot 23}\right]+
2\left[U_{\cdot 1}S_{\cdot 2}U_{\cdot 12}\right]\left[U_{\cdot 1}^{2}\right]
\nonumber
\\
&&
+2\left[U_{\cdot 1}S_{\cdot 1}\right]\left[{1},{2},{12}\right]_U\Bigg)
\nonumber
\\
&&
+g_{0}^{3}g_{1}^{2}\Bigg(U^{2}\left[S_{\cdot 1}S_{\cdot 2}U_{\cdot 13}U_{\cdot 23}\right]+
2U^{2}\left[U_{\cdot 1}S_{\cdot 2}U_{\cdot 23}
S_{\cdot 13}\right]+2U^{2}\left[U_{\cdot 1}S_{\cdot 2}U_{\cdot 13}S_{\cdot 23}\right]
\nonumber
\\
&&
+2U\left[{1}^{2}\right]_U \left[S_{\cdot 1}S_{\cdot 2}U_{\cdot 12}\right]+
4U\left[U_{\cdot 1}S_{\cdot 1}\right]\left[U_{\cdot 1}S_{\cdot 2}U_{\cdot 12}\right]
+2U\left[{1}, {2},{12}\right]_U\left[{1}^{2}\right]_S
\nonumber
\\
&&
+U^{2}\left[U_{\cdot 1}U_{\cdot 2}S_{\cdot 13}S_{\cdot 23}\right]+
2U\left[U_{\cdot 1}S_{\cdot 12}S_{\cdot 2}\right]\left[{1}^{2}\right]_U+2U\left[U_{\cdot 1}
U_{\cdot 2}S_{\cdot 12}\right]\left[U_{\cdot 1}S_{\cdot 1}\right]
\nonumber
\\
&&
+\left[{1}^{2}\right]_U^{2}\left[{1}^{2}\right]_S+3\left[U_{\cdot 1}S_{\cdot 1}\right]^{2}\left[{1}^{2}\right]_S\Bigg)
\nonumber
\\
&&
+2g_{0}g_{1}U\Bigg(U^{2}\left[S_{\cdot 1}S_{\cdot 2}U_{\cdot 13}S_{\cdot 23}\right]+
U\left[U_{\cdot 1}S_{\cdot 1}\right]\left[S_{\cdot 1}S_{\cdot 2}
U_{\cdot 12}\right]U\left[U_{\cdot 1}U_{\cdot 12}S_{\cdot 2}\right]\left[{1}^{2}\right]_S
\nonumber
\\
&&
+U^{2}\left[U_{\cdot 1}S_{\cdot 2}S_{\cdot 13}S_{\cdot 23}\right]+
U\left[{1}^{2}\right]_U\left[{1},S_{\cdot 2},{12}\right]_S
+2U\left[U_{\cdot 1}S_{\cdot 2}  S_{\cdot 12}\right]\left[U_{\cdot 1}S_{\cdot 1}\right]
\nonumber
\\
&&
+U\left[U_{\cdot 1}U_{\cdot 2}S_{\cdot 12}\right]\left[{1}^{2}\right]_S+
3\left[{1}^{2}\right]_U\left[U_{\cdot 1}S_{\cdot 1}\right]\left[{1}^{2}\right]_S+
\left[U_{\cdot 1}S_{\cdot 1}\right]^{3}\Bigg)
\nonumber
\\
&&
+2g_{0}^{3}g_{2}U\left[U_{\cdot 1}S_{\cdot 2}U_{\cdot 12}\right]\left[U_{\cdot 1}S_{\cdot 1}\right]
\nonumber
\\
&&
+2g_{0}^{2}g_{1}g_{2}\Bigg(U^{2}\left[U_{\cdot 1}S_{\cdot 1}\right]
\left[S_{\cdot 1}S_{\cdot 2}U_{\cdot 12}\right]+
U^{2}\left[U_{\cdot 1}S_{\cdot 2}
U_{\cdot 12}\right]\left[{1}^{2}\right]_S+U^{2}\left[U_{\cdot 1}S_{\cdot 2}S_{\cdot 12}\right]
\left[U_{\cdot 1}S_{\cdot 1}\right]
\nonumber
\\
&&
+U\left[{1}^{2}\right]_U\left[U_{\cdot 1}S_{\cdot 1}\right]\left[{1}^{2}\right]_S+
U\left[U_{\cdot 1}S_{\cdot 1}\right]^{3}\Bigg)
\nonumber
\\
&&
+2g_{0}g_{1}^{2}g_{2}U^{2}\Bigg(U\left[S_{\cdot 1}S_{\cdot 2}U_{\cdot 12}\right]
\left[{1}^{2}\right]_S+U\left[U_{\cdot 1}S_{\cdot 1}\right]
\left[{1},{2},{12}\right]_S+U\left[U_{\cdot 1}S_{\cdot 1}S_{\cdot 12}\right]\left[{1}^{2}\right]_S
\nonumber
\\
&&
+3\left[U_{\cdot 1}S_{\cdot 1}\right]^{2}\left[{1}^{2}\right]_S+\left[{1}^{2}\right]_U
\left[{1}^{2}\right]_S^{2}\Bigg)
\nonumber
\\
&&
+2g_{1}^{3}g_{2}U^{3}\Bigg(U\left[{1}^{2}\right]_S\left[{1},{2},{12}\right]_S+2\left[U_{\cdot 1}S_{\cdot 1}\right]
\left[{1}^{2}\right]_S^{2}\Bigg)
\nonumber
\\
&&
+g_{0}^{2}g_{2}^{2}U^{2}\left[U_{\cdot 1}S_{\cdot 1}\right]^{2}\left[{1}^{2}\right]_S
\nonumber
\\
&&
+2g_{0}g_{1}g_{2}^{2}U^{3}\left[U_{\cdot 1}S_{\cdot 1}\right]\left[{1}^{2}\right]_S^{2}
\nonumber
\\
&&
+g_{1}^{2}g_{2}^{2}U^{4}\left[{1}^{2}\right]_S^{3}\Bigg),
\nonumber
\end{eqnarray*}
\begin{eqnarray*}
\left[1,2,3,123\right]_T
&=&
g_{0}^{4}\left[{1},{2},{3},{123}\right]_U
\nonumber
\\
&&
+g_{0}^{3}g_{1}\Bigg(U\left[U_{\cdot 1}U_{\cdot 2}U_{\cdot 3}S_{\cdot 123}\right]+
3\left[{1}^{2}\right]_U\left[U_{\cdot 1}U_{\cdot 2}S_{\cdot 12}\right]
+3\left[{1},{2},{12}\right]_U\left[U_{\cdot 1}S_{\cdot 1}\right]
\nonumber
\\
&&
+3U\left[U_{\cdot 1}U_{\cdot 2}S_{\cdot 3}U_{\cdot 123}\right]\Bigg)
\nonumber
\\
&&
+g_{0}^{3}g_{2}\Bigg(3U\left[U_{\cdot 1}S_{\cdot 1}\right]\left[U_{\cdot 1}U_{\cdot 2}S_{\cdot 12}\right]+
3\left[{1}^{2}\right]_U\left[U_{\cdot 1}S_{\cdot 1}\right]^{2}\Bigg)
\nonumber
\\
&&
+g_{0}^{3}g_{3}U\left[U_{\cdot 1}S_{\cdot 1}\right]^{3}
\nonumber
\\
&&
+3g_{0}^{2}g_{1}^{2}U\Bigg(U\left[U_{\cdot 1}U_{\cdot 2}S_{\cdot 3}S_{\cdot 123}\right]+
2\left[{1}^{2}\right]_U\left[U_{\cdot 1}S_{\cdot 2}
S_{\cdot 12}\right]+\left[U_{\cdot 1}S_{\cdot 1}\right]\left[U_{\cdot 1}U_{\cdot 2}S_{\cdot 12}\right]
\nonumber
\\
&&
+ 2\left[U_{\cdot 1}S_{\cdot 1}\right]\left[U_{\cdot 1}S_{\cdot 2}U_{\cdot 12}\right]+
\left[{1},{2},{12}\right]_U\left[{1}^{2}\right]_S
+U\left[U_{\cdot 1}S_{\cdot 1}S_{\cdot 3}U_{\cdot 123}\right]\Bigg)
\nonumber
\\
&&
+ 3g_{0}^{2}g_{1}g_{2}U\Bigg(2U\left[U_{\cdot 1}S_{\cdot 1}\right]\left[U_{\cdot 1}S_{\cdot 2}S_{\cdot 12}\right]
+U\left[U_{\cdot 1}U_{\cdot 2}S_{\cdot 12}\right]  \left[{1}^{2}\right]_S
\nonumber
\\
&&
+2\left[{1}^{2}\right]_U\left[U_{\cdot 1}S_{\cdot 1}\right]\left[{1}^{2}\right]_S +
\left[U_{\cdot 1}S_{\cdot 1}\right]^{3}\Bigg)
\nonumber
\\
&&
+3g_{0}^{2}g_{1}g_{3}U^{2}\left[U_{\cdot 1}S_{\cdot 1}\right]^{2}\left[{1}^{2}\right]_S
\nonumber
\\
&&
+g_{0}g_{1}^{3}U^{2}\Bigg(3U\left[U_{\cdot 1}S_{\cdot 2}S_{\cdot 3}S_{\cdot 123}\right]
+3\left[{1}^{2}\right]_U\left[S_{\cdot 1}S_{\cdot 2}S_{\cdot 12}\right]
+6\left[U_{\cdot 1}S_{\cdot 1}\right]\left[U_{\cdot 1}S_{\cdot 2}S_{\cdot 12}\right]
\nonumber
\\
&&
+3\left[U_{\cdot 1}S_{\cdot 1}\right]\left[S_{\cdot 1}S_{\cdot 2}U_{\cdot 12}\right]+
6\left[U_{\cdot 1}S_{\cdot 2}U_{\cdot 12}\right]\left[S_{\cdot 1}^{2}\right]+U\left[S_{\cdot 1}
 S_{\cdot 2}S_{\cdot 3}U_{\cdot 123}\right]\Bigg)
\nonumber
\\
&&
+3g_{0}g_{1}^{2}g_{2}U^{2}\Bigg(U\left[U_{\cdot 1}S_{\cdot 1}\right]\left[{1},{2},{12}\right]_S+
2U\left[U_{\cdot 1}S_{\cdot 2}
S_{\cdot 12}\right]\left[{1}^{2}\right]_S
\nonumber
\\
&&
+\left[{1}^{2}\right]_U\left[{1}^{2}\right]_S^{2}+
2\left[U_{\cdot 1}S_{\cdot 1}\right]^{2}\left[{1}^{2}\right]_S\Bigg)
\nonumber
\\
&&
+3g_{0}g_{1}^{2}g_{3}U^{3}\left[U_{\cdot 1}S_{\cdot 1}\right]\left[{1}^{2}\right]_S^{2}
\nonumber
\\
&&
+g_{1}^{4}U^{3}\Bigg(U\left[{1},{2},{3},{123}\right]_S+
3\left[U_{\cdot 1}S_{\cdot 1}\right]\left[{1},{2},{12}\right]_S
+3\left[S_{\cdot 1}S_{\cdot 2}U_{\cdot 12}\right]\left[{1}^{2}\right]_S\Bigg)
\nonumber
\\
&&
+3g_{1}^{3}g_{2}U^{3}\Bigg(U\left[{1}\right]_S^{2}\left[{1},{2},{12}\right]_S+
\left[U_{\cdot 1}S_{\cdot 1}\right]\left[{1}^{2}\right]_S^{2}\Bigg)
\nonumber
\\
&&
+g_{1}^{3}g_{3}U^{4}\left[{1}^{2}\right]_S^{3}.
\nonumber
\end{eqnarray*}

\section*{Appendix B}
\renewcommand{\theequation}{$\mbox{B.\arabic{equation}}$}
\setcounter{equation}{0}

Here, we give bracket functions for moments to order seven, that is expressions for
\begin{eqnarray*}
\left( 1_r^a 11_s^b 2_t^2 12_u^d\cdots \right) =
\int_{-\infty}^\infty \cdots \int_{-\infty}^\infty \mu_{r \cdot 1}^a \mu_{s \cdot 11}^b
\mu_{t \cdot 2}^c \mu_{u \cdot 12}^d \cdots dF_1 \left( x_1 \right) dF_2 \left( x_2 \right) \cdots
\end{eqnarray*}
up to $N=ar+bs+ct+\cdots =7$, where now $\mu_1$ means $\mu$.
This enables one to obtain the Edgeworth-Cornish-Fisher expansions and bias
reduction for any smooth function of $(\mu,\mu_2,\mu_3,\ldots)$.
We exclude separable terms like $(1^2_a 2^2_b)=(1^2_a)(1^2_b)$.
For each $N$, we order the terms by its partition functions.
\begin{eqnarray*}
N &=& 2 \ \left(2 \mbox{ terms}\right):
\\
1^2:&& \ \left(1_1^2\right)=\mu_2,
\\
2: && \ \left(11_2\right)=-2\mu_2.
\\
N &=& 3 \ \left(5 \mbox{ terms}\right):
\\
1^3: && \ \left(1_1^3\right)=\mu_3,
\\
12: &&  \ \left(1_1 1_2\right)=\mu_3,
\
\left(1_1 11_2\right)=-2\mu_3,
\\
3:  && \ \left(11_3\right)=-6\mu_3,
\
\left(111_3\right)=-12\mu_3.
\\
N &=& 4 \ \left(13 \mbox{ terms}\right):
\\
1^4: && \ \left(1_1^4\right)=\mu_4,
\\
1^2 2: && \ \left(1_1^2 1_2\right)=\mu_4-\mu_2^2,
\
\left(1_1^2 11_2\right)=-2\mu_4,
\\
13: && \ \left(1_1 1_3\right)=\mu_4-\mu_2^2,
\
\left(1_1 11_3\right)=-6\left(\mu_4-\mu_2^2\right),
\
\left(1_1 111_3\right)=12\mu_4,
\\
&&   \ \left(1_1 122_3\right)=12\mu_2^2,
\\
2^2: &&
\ \left(1_2^2\right)=\mu_4-\mu_2^2,
\
\left(1_2 11_2\right)=-2\left(\mu_4-\mu_2^2\right),
\
\left(11_2^2\right)=4\mu_4,
\
\left(12_2^2\right)=4\mu_2^2,
\\
4: && \left(1111_4\right)=-72\mu_4,
\
\left(1122_4\right)=-72\mu_2^2.
\\
N &=& 5 \ \left(43 \mbox{ terms}\right):
\\
1^5: && \ \left(1_1^5\right)=\mu_5,
\\
1^3 2: && \ \left(1_1^3 1_2\right)=\mu_5-\mu_3\mu_2,
\
\left(1_1^3 11_2\right)=\mu_5-2\mu_3\mu_2,
\
\left(1_1^2 2_1 12_2\right)=-2\mu_3\mu_2,
\\
1^2 3: && \ \left(1_1^2 1_3\right)=\mu_3\mu_2,
\
\left(1_1^2 11_3\right)=-6\left(\mu_5-\mu_3\mu_2\right),
\
\left(1_1^2 111_3\right)=12\mu_5,
\\
&&   \ \left(1_1^2 122_3\right)=12\mu_3\mu_2,
\
\left(1_1 2_1 12_3\right)=-2\mu_2^3,
\
\left(1_1 2_1 112_3\right)=12\mu_3\mu_2,
\\
1 2^2: && \ \left(1_1 1_2^2\right)=\mu_5-2\mu_3\mu_2,
\
\left(1_1 1_2 11_2\right)=-2\left(\mu_5-\mu_3\mu_2\right),
\
\left(1_1 2_2 12_2\right)=4\mu_3\mu_2,
\\
&&   \ \left(1_1 11_2^2\right)=4\mu_5,
\
\left(1_1 12_2^2\right)=4\mu_3\mu_2,
\\
14: && \ \left(1_1 1_4\right)=\mu_5-4\mu_3\mu_2,
\
\left(1_1 11_4\right)=-8\mu_5+20\mu_3\mu_2,
\\
&&   \ \left(1_1 111_4\right)=36\left(\mu_5-\mu_3\mu_2\right),
\
\left(1_1 1111_4\right)=-72\mu_5,
\\
&&   \ \left(1_1 122_4\right)=6\mu_3\mu_2,
\
\left(1_1 1122_4\right)=-72\mu_3\mu_2,
\\
&&   \ \left(1_1 1222_4\right)=-72\mu_3\mu_2,
\\
23: && \ \left(1_2 1_3\right)=\mu_5-4\mu_3\mu_2,
\
\left(1_2 11_3\right)=-6\left(\mu_5-\mu_3\mu_2\right),
\
\left(1_2 111_3\right)=12\left(\mu_5-\mu_3\mu_2\right),
\\
&&   \ \left(11_2 1_3\right)=-2\left(\mu_5-4\mu_3\mu_2\right),
\
\left(11_2 11_3\right)=12\left(\mu_5-\mu_3\mu_2\right),
\
\left(11_2 111_3\right)=-24\mu_5,
\\
&&   \ \left(11_2 122_3\right)=-24\mu_3\mu_2,
\
\left(12_2 112_3\right)=-24\mu_3\mu_2,
\\
5:&& \ \left(11_5\right)= -10\left(\mu_5-2\mu_3\mu_2\right),
\
\left(111_5\right)= 60\left(\mu_5-2\mu_3\mu_2\right),
\
\left(1111_5\right)= -240\left(\mu_5-\mu_3\mu_2\right),
\\
&&  \ \left(11111_5\right)= 480\mu_5,
\
\left(1122_5\right)= 0,
\
\left(11122_5\right)= 480\mu_3\mu_2.
\end{eqnarray*}
\begin{eqnarray*}
N &=& 6 \ \left(85 \mbox{ terms}\right):
\\
1^6: && \ \left(1_1^6\right)=\mu_6,
\\
1^4 2: && \ \left(1_1^4 1_2\right)=\mu_6-\mu_4\mu_2,
\
\left(1_1^4 11_2\right)=-2\mu_6,
\
\left(1_1^3 2_1 12_2\right)=-2\mu_4\mu_2,
\\
&&  \ \left(1_1^2 2_1^2 12_2\right)=-2\mu_3^2,
\\
1^3 3: && \ \left(1_1^3 1_3\right)=\mu_6-3\mu_4\mu_2-\mu_3^2,
\
\left(1_1^3 11_3\right)=-6\left(\mu_6-\mu_4\mu_2\right),
\\
&&  \ \left(1_1^3 111_3\right)=12\mu_6,
\
\left(1_1^3 122_3\right)=12\mu_4\mu_2,
\
\left(1_1^2 2_1 12_3\right)=-3\left(\mu_4\mu_2+\mu_3^2-\mu_2^3\right),
\\
&&  \  \left(1_1^2 2_1 112_3\right)=12\mu_4\mu_2,
\
\left(1_1^2 2_1 122_3\right)=12\mu_3^2,
\
\left(1_1 2_1 3_1 123_3\right)=12\mu_2^3,
\\
1^2 4:&& \ \left(1_1^2 1_4\right)=\mu_6-4\mu_4\mu_2-4\mu_3^2,
\
\left(1_1^2 11_4\right)=-4\left(2\mu_6-3\mu_4\mu_2-2\mu_3^2\right),
\\
&&   \ \left(1_1^2 111_4\right)=-12\left(\mu_6-\mu_4\mu_2\right),
\
\left(1_1^2 1111_4\right)=-72\mu_6,
\\
&&   \ \left(1_1^2 122_4\right)=-12\left(\mu_4\mu_2+2\mu_3^2-\mu_2^3\right),
\
\left(1_1^2 1122_4\right)=-72\mu_4\mu_2,
\\
&&   \ \left(1_1^2 1222_4\right)=-72\mu_3^2,
\
\left(1_1 2_1 12_4\right)=-4\left(2\mu_4\mu_2-3\mu_2^3\right),
\\
&&   \ \left(1_1 2_1 112_4\right)=-12\left(2\mu_4\mu_2+\mu_3^2-2\mu_2^3\right),
\
\left(1_1 2_1 1112_4\right)=480\mu_4\mu_2,
\\
&&   \ \left(1_1 2_1 1122_4\right)=480\mu_3^2,
\
\left(1_1 2_1 1233_4\right)=480\mu_2^3,
\\
1^2 2^2:&& \ \left(1_1^2 11_2^2\right)=4\mu_6,
\
\left(1_1^2 12_2^2\right)=4\mu_4\mu_2,
\
\left(1_1 2_1 11_2 22_2\right)=4\mu_3^2,
\\
&&   \ \left(1_1 2_1 12_2^2\right)=-2\mu_3^2,
\\
123:&& \ \left(1_1 11_2 111_3\right)=-24\mu_6,
\
\left(1_1 11_2 122_3\right)=-24\mu_4\mu_2,
\\
&&  \ \left(1_1 12_2 112_3\right)=-24\mu_4\mu_2,
\
\left(1_1 12_2 122_3\right)=-24\mu_3^2,
\\
&&  \ \left(1_1 12_2 222_3\right)=-24\mu_4\mu_2,
\\
15:&& \ \left(1_1 1_5\right)= \mu_6-5\mu_4\mu_2,
\
\left(1_1 11_5\right)= -10\left(\mu_6-\mu_4\mu_2-2\mu_3^2\right),
\\
&&  \ \left(1_1 111_5\right)= 60\left(\mu_6-\mu_4\mu_2-\mu_3^2\right),
\
\left(1_1 122_5\right)= 60\mu_4\mu_2,
\\
&&  \ \left(1_1 1111_5\right)= 240\left(\mu_6-\mu_4\mu_2\right),
\
\left(1_1 1122_5\right)= 120\left(\mu_4\mu_2+\mu_3^2-\mu_2^3\right),
\\
&&  \ \left(1_1 1222_5\right)= 60\left(3\mu_4\mu_2+\mu_3^2-3\mu_2^3\right),
\
\left(1_1 11111_5\right)= 480\mu_6,
\\
&&  \ \left(1_1 11122_5\right)= 480\mu_4\mu_2,
\
\left(1_1 11222_5\right)= 480\mu_3^2,
\\
&&  \ \left(1_1 12222_5\right)= 480\mu_4\mu_2,
\
\left(1_1 12233_5\right)= 480\mu_2^3,
\\
2^3:
&& \ \left(1_2^3\right)=\mu_6-3\mu_4\mu_2+2\mu_2^3,
\
\left(1_2 1_2 11_2\right)=-2\left(\mu_6-2\mu_4\mu_2+\mu_2^3\right),
\\
&&  \ \left(1_2 2_2 1_2\right)=2\mu_2^3,
\
\left(1_2  11_2^2\right)=4\left(\mu_6-\mu_4\mu_2\right),
\\
&&  \ \left(1_2  12_2^2\right)=4\left(\mu_4\mu_2-\mu_2^3\right),
\
\left(11_2^3\right)=-8\mu_6,
\\
&&  \ \left(11_2  12_2^2\right)=-8\mu_4\mu_2,
\
\left(12_2 23_2 31_2\right)=-8\mu_2^3,
\\
24:
&& \ \left(1_2 1_4\right)=\mu_6-\mu_4\mu_2-4\mu_2^3,
\
\left(1_2 11_4\right)=-4\left(2\mu_6-3\mu_4\mu_2-2\mu_3^2+3\mu_2^3\right),
\\
&&  \ \left(1_2 111_4\right)=-36\left(\mu_6-2\mu_4\mu_2+2\mu_2^3\right),
\
\left(1_2 1111_4\right)=-72\left(\mu_6-\mu_4\mu_2\right),
\\
&&  \ \left(1_2 1122_4\right)=-72\left(\mu_4\mu_2-\mu_2^3\right),
\
\left(11_2 1_4\right)=-2\left(\mu_6-\mu_4\mu_2-4\mu_2^3\right),
\\
&&  \ \left(11_2 11_4\right)=8\left(2\mu_6-3\mu_4\mu_2-2\mu_3^2\right),
\
\left(11_2 111_4\right)=72\left(\mu_6-\mu_4\mu_2\right),
\\
&&  \ \left(11_2 1111_4\right)=144\mu_6,
\
\left(11_2 1122_4\right)=144\mu_4\mu_2,
\\
&&  \ \left(12_2 12_4\right)=8\left(2\mu_4\mu_2-3\mu_2^3\right),
\
\left(12_2 112_4\right)=-4\left(2\mu_4\mu_2+\mu_3^2-2\mu_2^3\right),
\\
&&  \ \left(12_2 1112_4\right)=
\
\left(12_2 1122_4\right)=144\mu_3^2,
\\
3^2:
&& \ \left(1_3^2\right)=\mu_6-6\mu_4\mu_2-\mu_3^2+9\mu_2^3,
\
\left(1_3 11_3\right)=-6\left(\mu_6-4\mu_4\mu_2-\mu_3^2+3\mu_2^3\right),
\\
&&  \ \left(1_3 111_3\right)=12\left(\mu_6-3\mu_4\mu_2-\mu_3^2\right),
\
\left(1_3 122_3\right)=12\left(\mu_4\mu_2-3\mu_2^3\right),
\\
&&  \ \left(11_3 122_3\right)=-72\left(\mu_4\mu_2-\mu_2^3\right),
\
\left(12_3^2\right)=18\left(\mu_4\mu_2+\mu_3^2-\mu_2^3\right),
\\
&& \left(12_3 112_3\right)=-36\left(\mu_4\mu_2+\mu_3^2-\mu_2^3\right),
\end{eqnarray*}
\begin{eqnarray*}
6:
&& \ \left(11_6\right)=-6\left(2\mu_6-5\mu_4\mu_2\right),
\
\left(111_6\right)=30\left(3\mu_6-3\mu_4\mu_2-4\mu_3^2\right),
\\
&&  \ \left(1111_6\right)=-120\left(\mu_6-\mu_4\mu_2+\mu_3^2\right),
\
\left(11111_6\right)=-1800\left(\mu_6-\mu_4\mu_2\right),
\\
&&  \ \left(1122_6\right)=-120\left(4\mu_4\mu_2-3\mu_2^3\right),
\
\left(11122_6\right)=-360\left(3\mu_4\mu_2+2\mu_3^2-3\mu_2^3\right),
\\
&&  \ \left(111111_6\right)=-3600\mu_6,
\
\left(111122_6\right)=-3600\mu_4\mu_2,
\\
&&  \ \left(111222_6\right)=-3600\mu_3^2,
\
\left(112233_6\right)=-3600\mu_2^3.
\end{eqnarray*}
As noted for $N=2, \ldots, 6$ there are 2,5,13,43 and 85 terms, or without $\mu$, 1,2,6,20 and 39 terms.
We now give the 90 terms for $N=7$ without $\mu$.
\begin{eqnarray*}
N &=& 7:
\\
2^2 3:
&&
\ \left(1_2^2 1_3\right)=\mu_5-4\mu_3\mu_2,
\\
&&
\ \left(1_2^2 11_3\right)=-6\left(\mu_7-3\mu_5\mu_2+3\mu_3\mu_2^2\right),
\\
&&
\ \left(1_2^2 111_3\right)=-24\mu_5,
\\
&&
\ \left(1_2^2 122_3\right)= \ \left(1_2 2_2 122_3\right)=-24\mu_3\mu_2,
\\
&&
\ \left(1_2 11_2 1_3\right)=2\left(3\mu_5\mu_2-\mu_4\mu_3+\mu_3^2\right),
\\
&&
\ \left(1_2 11_2 11_3\right)=12\left(\mu_7-2\mu_5\mu_2+\mu_3\mu_2^2\right),
\\
&&
\ \left(1_2 11_2 111_3\right)=-24\left(\mu_7-\mu_5\mu_2\right),
\\
&&
\ \left(1_2 11_2 122_3\right)=-12\left(\mu_5-\mu_3\mu_2\right)\mu_2,
\\
&&
\ \left(1_2 12_2 2_3\right)=4\left(\mu_4-\mu_2^2\right)\mu_3,
\\
&&
\ \left(1_2 12_2 12_3\right)=6\left(\mu_5\mu_2-2\mu_4\mu_3\right),
\\
&&
\ \left(1_2 12_2 112_3\right)=-24\left(\mu_5-\mu_3\mu_2\right)\mu_2,
\\
&&
\ \left(1_2 12_2 122_3\right)=-24\left(\mu_4-\mu_2^2\right)\mu_3,
\\
&&
\ \left(1_2 12_2 222_3\right)=-24\mu_4\mu_3,
\\
&&
\ \left(1_2 12_2 233_3\right)=-24\mu_3\mu_2^2,
\\
25:
&&
\ \left(1_2 1_5\right)=\mu_7-\mu_5\mu_2-5\mu_4\mu_3,
\\
&&
\ \left(1_2 11_5\right)=-10\left(\mu_7-\mu_5\mu_2+3\mu_4\mu_3-2\mu_3\mu_2^2\right),
\\
&&
\ \left(1_2 111_5\right)=60\left(\mu_7-2\mu_5\mu_2-\mu_4\mu_3+2\mu_3\mu_2^2\right),
\\
&&
\ \left(1_2 122_5\right)=20\left(\mu_5\mu_2+2\mu_4\mu_3-4\mu_3\mu_2^2\right),
\\
&&
\ \left(1_2 1111_5\right)=\ \left(1_2 11111_5\right)=480\left(\mu_7-\mu_5\mu_2\right),
\\
&&
\ \left(1_2 1122_5\right)=120\left(\mu_5\mu_2+\mu_4\mu_3-3\mu_3\mu_2^2\right),
\\
&&
\ \left(1_2 11122_5\right) =\ \left(11_2 11122_5\right)=480\left(\mu_5-\mu_3\mu_2\right)\mu_2,
\\
&&
\ \left(1_2 11222_5\right) = \ \left(11_2 11222_5\right) =\ -\left(12_2 1122_5\right)= 480\left(\mu_4-\mu_2^2\right)\mu_3,
\\
&&
\ \left(1_2 12222_5\right)=\ \left(11_2 12222_5\right)= 480\mu_4\mu_3,
\\
&&
\ \left(1_2 12233_5\right) = \ \left(11_2 12233_5\right) =480\mu_3\mu_2^2,
\\
&&
\ \left(11_2 1_5\right)=-2\left(\mu_7-\mu_5\mu_2-5\mu_4\mu_3\right),
\\
&&
\ \left(11_2 11_5\right)=20\left(\mu_7-3\mu_4\mu_3\right),
\\
&&
\ \left(11_2 111_5\right)=-120\left(\mu_7-\mu_5\mu_2-\mu_4\mu_3\right),
\\
&&
\ \left(11_2 122_5\right)=-40\left(\mu_5\mu_2+2\mu_4\mu_3-4\mu_3\mu_2^2\right),
\\
&&
\ \left(11_2 1111_5\right)=240\left(\mu_7-2\mu_5\mu_2+\mu_3\mu_2^2\right),
\\
&&
\ \left(11_2 1122_5\right)=120\left(\mu_5\mu_2+\mu_4\mu_3-3\mu_3\mu_2^2\right),
\\
&&
\ \left(11_2 11111_5\right)=480\left(\mu_7-\mu_5\mu_2\right),
\\
&&
\ \left(12_2 112_5\right)=-40\left(2\mu_5\mu_2+\mu_4\mu_3-5\mu_3\mu_2^2\right),
\\
&&
\ \left(12_2 1112_5\right)=-120\left(3\mu_5\mu_2+\mu_4\mu_3-3\mu_3\mu_2^2\right),
\\
&&
\ \left(12_2 1222_5\right)=-120\left(3\mu_5\mu_2+\mu_4\mu_3-3\mu_3\mu_2^2\right),
\\
&&
\ \left(12_2 11112_5\right)=-960\mu_5\mu_2,
\\
&&
\ \left(12_2 11122_5\right)=-960\mu_4\mu_3,
\\
&&
\ \left(12_2 12233_5\right)=-960\mu_3\mu_2^2,
\end{eqnarray*}
\begin{eqnarray*}
34: && \left(1_3 1_4\right) =\mu_7-3\mu_5\mu_2-5\mu_4\mu_3+12\mu_3\mu_2^2,
\\
&&
\left(1_3 11_4\right) =-4\left(2\mu_7-9\mu_5\mu_2-4\mu_4\mu_3-18\mu_3\mu_2^2\right),
\\
&&
\left(1_3 111_4\right) =-36\left(\mu_7-4\mu_5\mu_2-\mu_4\mu_3+4\mu_3\mu_2^2\right),
\\
&&
\left(1_3 1111_4\right) =72\left(\mu_7-3\mu_5\mu_2-\mu_4\mu_3\right),
\\
&&
\left(1_3 1122_4\right) =72\left(\mu_5-4\mu_3\mu_2\right)\mu_2,
\\
&&
\left(11_3 1_4\right) = \ \left(11_3 11_4\right)/4 =-6\left(\mu_7-\mu_5\mu_2-4\mu_4\mu_3+4\mu_3\mu_2^2\right),
\\
&&
\left(11_3 111_4\right) =216\left(\mu_7-2\mu_5\mu_2+\mu_3\mu_2^2\right),
\\
&&
\left(11_3 1111_4\right) =432\left(\mu_7-\mu_5\mu_2\right),
\\
&&
\left(11_3 1122_4\right) =432\left(\mu_5-\mu_3\mu_2\right)\mu_2,
\\
&&
\left(11_3 11222_4\right) =432\left(\mu_4-\mu_2^2\right)\mu_3,
\\
&&
\left(12_3 12_4\right) =24\left(\mu_5\mu_2+\mu_4\mu_3-5\mu_3\mu_2^2\right),
\\
&&
\left(12_3 112_4\right) =24\left(\mu_5\mu_2+\mu_4\mu_3-4\mu_3\mu_2^2\right),
\\
&&
\left(12_3 1112_4\right) =\ \left(12_3 1222_4\right) =216\left(\mu_5\mu_2+\mu_4\mu_3-\mu_3\mu_2^2\right),
\\
&&
\left(12_3 1122_4\right) =432\left(\mu_4-\mu_2^2\right)\mu_3,
\\
&&
\left(12_3 1233_4\right) =432\mu_3\mu_2^2,
\\
7:
&&
\left(11_7\right) = -14\left(\mu_7-3\mu_5\mu_2\right),
\\
&&
\left(111_7\right) = 42\left(3\mu_7-3\mu_5\mu_2-5\mu_4\mu_3\right),
\\
&&
\left(1111_7\right) = -840\left(\mu_7-2\mu_4\mu_3\right),
\\
&&
\left(1122_7\right) = -140\left(4\mu_5\mu_2-3\mu_3\mu_2^2\right),
\\
&&
\left(11111_7\right) = 840\left(5\mu_7-3\mu_5\mu_2-5\mu_4\mu_3\right),
\\
&&
\left(11122_7\right) = 420\left(3\mu_5\mu_2+2\mu_4\mu_3-6\mu_3\mu_2^2\right),
\\
&&
\left(111 111_7\right) = 15120\left(\mu_7-\mu_5\mu_2\right),
\\
&&
\left(111 122_7\right) = 7!\left(2\mu_5\mu_2+\mu_4\mu_3-2\mu_3\mu_2^2\right),
\\
&&
\left(111 222_7\right) = 15120\left(\mu_4-\mu_2^2\right)\mu_3,
\\
&&
\left(11 22 33_7\right) = 15120\mu_3\mu_2^2,
\\
&&
\left(111 1111_7\right) = 30240\mu_7,
\\
&&
\left(111 1122_7\right) = 30240\mu_5\mu_2,
\\
&&
\left(111 1222_7\right) = 30240\mu_4\mu_3,
\\
&&
\left(111 22 33_7\right) = 30240\mu_3\mu_2^2.
\end{eqnarray*}
For $F$ symmetric terms corresponding to odd $N$ reduce to zero.

\newpage

\centerline{\epsfig{file=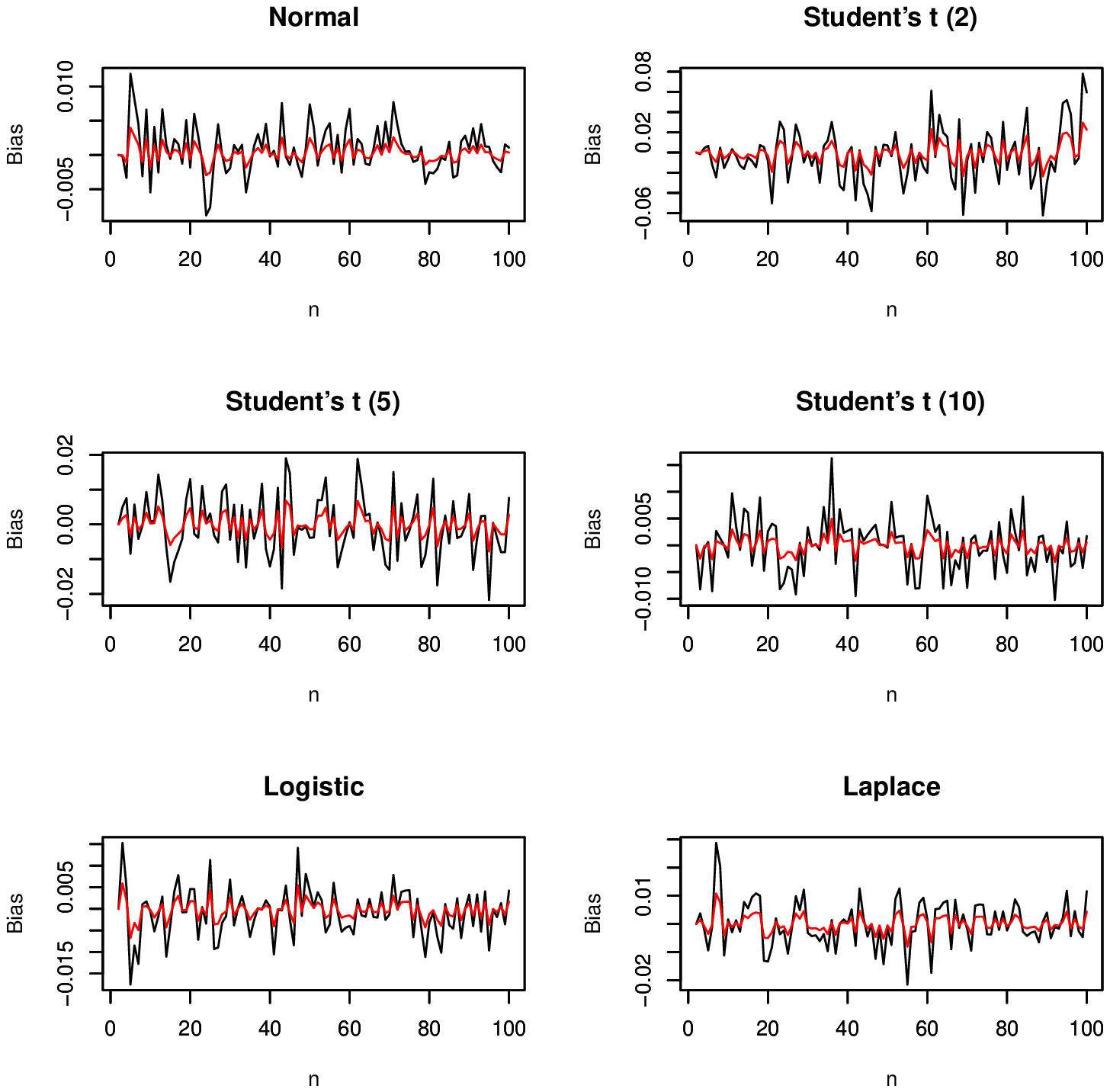,width=6in,height=8in}}
\noindent
{\bf Figure 4.1}~~Biases of the usual (black) and bias reduced (red) estimators
of skewness versus $n = 2, 3, \ldots, 100$.

\end{document}